\documentclass[twocolumn]{aastex631}
\hypersetup{linkcolor=red,citecolor=blue,filecolor=cyan,urlcolor=magenta}

\begin{document}
\title{Two Epochs of VLBI Observations of 8 KISSR Seyfert \& LINER Galaxies: Suggestions of Fast and Filamentary Outflows} 
\correspondingauthor{Preeti Kharb}
\email{kharb@ncra.tifr.res.in}
\author[0000-0003-3203-1613]{Preeti Kharb}
\affiliation{National Centre for Radio Astrophysics (NCRA) - Tata Institute of Fundamental Research (TIFR), \\
S. P. Pune University Campus, Ganeshkhind, Pune 411007, Maharashtra, India}
\author[0000-0001-9707-3895]{Anderson Caproni}
 \affiliation{N\'ucleo de Astrof\'\i sica, Universidade Cidade de S\~ao Paulo, R. Galv\~ao Bueno 868, Liberdade, S\~ao Paulo, SP, 01506-000, Brazil}
\author[0009-0000-1447-5419]{Salmoli Ghosh}
\affiliation{National Centre for Radio Astrophysics (NCRA) - Tata Institute of Fundamental Research (TIFR), \\
S. P. Pune University Campus, Ganeshkhind, Pune 411007, Maharashtra, India}
\author[0000-0001-8252-4753]{Daniel A. Schwartz}
\affiliation{Center for Astrophysics $\vert$ Harvard \& Smithsonian, 60 Garden St, Cambridge, MA 02138, USA}
\author[0000-0001-8996-6474]{Mousumi Das}
\affiliation{Indian Institute of Astrophysics, 2nd Block, Koramangala, Bangalore 560034, India}
\author[0000-0002-5331-6098]{Smitha Subramanian}
\affiliation{Indian Institute of Astrophysics, 2nd Block, Koramangala, Bangalore 560034, India}
\author[0000-0003-3295-6595]{Sravani Vaddi}
\affiliation{Arecibo Observatory: Arecibo, Puerto Rico, USA}

\begin{abstract}
We present here the results from a second epoch of phase-referenced VLBA observations of 8 Seyfert and LINER galaxies from the KISSR sample. These sources were chosen based on the presence of double peaks or asymmetries in their emission lines as observed in SDSS spectra. Parsec-scale radio emission is detected in 7 of the 8 sources in the second epoch. Jet-like features appear to persist over a $\sim4-9$ year timeline in these `radio-quiet' AGN. A few sources like KISSR1494, however, show significantly different structures after a 9~year interval. KISSR102, which was previously suggested to be a binary black hole candidate based on the presence of two compact cores, shows the tentative signatures of superluminal jet motion ($1.05\pm0.45$c). Tentative superluminal motion in a jet knot has been reported in another source, KISSR872 ($1.65\pm0.57$c). We present 1.5~GHz images from the VLA A-array of 4 sources. These images reveal core-lobe or core-halo structures. The alignment of the VLBI jet direction with the kpc-scale spectral index gradient, as well as the mismatch between the star formation rate derived from the radio and H$\alpha$ line emission, support the suggestion that the kpc-scale emission is AGN-jet-related. The jets in KISSR sources appear to lose collimation over spatial scales between 200~parsec and 1 kpc. Overall, the characteristics of the KISSR jets are reminiscent of similar properties observed in VLBI monitoring studies of `radio-loud' AGN jets even as subtle differences related to the compactness and brightness of jet features remain.
\end{abstract}
\keywords{Seyfert galaxies --- LINER galaxies --- Very long baseline interferometry --- Jets}

\section{Introduction} \label{sec:intro}
Active galactic nuclei (AGN) are the luminous centres of galaxies that are powered by the release of gravitational potential energy as matter accretes onto supermassive black holes (SMBHs, masses $\sim10^6-10^9$~M$_\odot$) through accretion disks \citep{Rees1984,Peterson97}. Seyfert and low-ionization nuclear emission-line region (LINER) galaxies make the vast majority of the AGN population \citep{Ho2008,Heckman2014}. AGN showing only narrow emission lines in their spectra are classified as type 2, while those showing both broad and narrow lines are classified as type 1 \citep{Khachikian1974}. It has been postulated that types 1 and 2 differ only by orientation with broad line emission being obscured due to dusty tori \citep{Osterbrock1978,Antonucci1985,Antonucci93}. The half opening angle of the obscuring tori may be $\sim50^\circ$ based on the space density of Seyfert type 1s and 2s, emission-line bicone images or X-ray monitoring studies \citep[e.g.,][]{Woltjer1990, Schmitt2001, Simpson96, Miniutti2014}.

The presence of broad and/or narrow emission lines in the spectra is one of the hallmarks of an AGN. A small fraction of AGN \citep[$\sim1\%$ in the SDSS\footnote{Sloan Digital Sky Survey \citep{York00}} survey;][]{Wang09,Liu2010} show double-peaked emission lines in their spectra. These are referred to as double-peaked AGN (DPAGN). In this paper, we look only at AGN that show double peaks in `narrow' emission lines, whose presence has been suggested to arise either due to the presence of binary SMBHs (predicted in hierarchical galaxy evolution models) with individual associated narrow-line regions or NLRs \citep{Zhou2004,Gerke2007} disk-like NLRs \citep{Smith2012,Nevin2016, Zhang2025}, NLR kinematics \citep{Shen2011,Fu2012}, gas outflows \citep{Fischer2011,Comerford2018}, jet-NLR interaction \citep{Whittle1988,Rosario2010}, or a combination of these and other factors \citep[][]{Ge2012, Muller2015, Maschmann2023}.

While Seyferts and LINER galaxies have been categorised as `radio-quiet' (RQ) AGN \citep{Kellermann89,Ho2008}, very long baseline interferometry (VLBI) observations of Seyfert and LINER galaxies have revealed the presence of weak radio cores and jets in them \citep{Falcke2000,Ulvestad2001, Doi2013,Orienti10,Baldi2018,Kharb2021}. Similarly on arcsecond-scales, 100 pc-scale and kpc-scale radio structures have been detected as well \citep{Baum93,Wilson1995,Colbert1996,Gallimore06,Kharb06,Rao2023}. While the existence of a clear `radio-loud'/`radio-quiet' AGN divide has been debated \citep[e.g.,][]{White2000,Cirasuolo2003}, and several definitions have been put forth to classify AGN into these RL/RQ AGN categories \citep[e.g.,][]{Terashima2003,Kellermann2016,Klindt2019}, it is fair to state that RQ AGN have radio jets that typically do not extend beyond the confines of their host galaxies, unlike `radio-loud' AGN.

To understand the nature of the radio outflows in RQ AGN and the origin of double-peaked emission lines, we identified an initial sample of 9 type~2 Seyfert and LINER galaxies belonging to the larger KPNO Internal Spectroscopic Red (KISSR) Survey of spiral or disk emission-line galaxies \citep{Wegner03}. These 9 sources were selected based on the presence of either double peaks or asymmetries in one or more of the narrow emission lines of [S {\sc ii}], H$\alpha$, [N {\sc ii}], or H$\beta$ in SDSS spectra. Sources showing asymmetry only in the [O {\sc iii}] lines were not selected as such asymmetries could be a signature of gas outflows. Additionally, for the feasibility of a VLBI study, a detection in the VLA FIRST\footnote{Faint Images of the Radio Sky at Twenty-Centimeters \citep{Becker1995}} survey was a requirement. All sources are unresolved in VLA FIRST images (with the FIRST resolution of $\theta\sim5\arcsec$ translating to spatial scales of $\sim5-6$~kpc in these sources) and had 1.4~GHz radio flux densities ranging from 2.5 to 23.0 mJy. Dual-frequency phase-referenced VLBA observations of these 9 DPAGN were carried out from 2013 to 2019 and their results have been presented in \citet{Kharb2015,Kharb2017a,Kharb2019,Kharb2020,Kharb2021}. We briefly discuss the main results below.

\begin{deluxetable*}{lcclcclccl}
\tabletypesize{\small}
\tablewidth{0pt} 
\tablecaption{{VLBA Observation Details \label{tab:obs}}}
\tablehead{
\colhead{Source} & \colhead{RA} & \colhead{DEC} & \colhead{Date}& \colhead{Sun}& \colhead{Frequency} & \colhead{PhaseCal} & \colhead{Separation} & \colhead{No. of} & \colhead{Epoch-I}\\
\colhead{name} & \colhead{h~m~s} & \colhead{$\degr~\arcmin~\arcsec$} & \colhead{D/M/Y}&  \colhead{ deg}&\colhead{GHz}& \colhead{name} & \colhead{deg} & \colhead{antennas} & \colhead{Ref}} 
\colnumbers
\startdata
{KISSR102}& 12 41 35.143 & 28 50 36.270 & 15/09/2022& 30.7 & 1.54 & 1246+285 & 1.78 & 8  & \citet{Kharb2020}\\
{}&              &                      & 18/09/2022& 30.4 & 4.98 & 1246+285 &  & 7 & \\
{KISSR434}& 14 03 45.022 & 29 21 43.980 & 29/11/2022& 60.2 & 1.54 & 1404+286 & 1.15 & 8 & \citet{Kharb2019}\\
{}&              &                      & 01/09/2022& 52.1 & 4.98  & 1404+286 & & 9 & \\
{KISSR618}& 15 02 28.776 & 28 58 15.600        & 04/09/2022& 62.9 & 1.54 & J1454+29 & 1.98 & 9 & \citet{Kharb2021}\\
{KISSR872$^{\ast}$}& 15 50 09.805 & 29 11 07.230 & 13/09/2022& 67.9 & 1.54 & 1537+279 & 2.72 & 8 & \citet{Kharb2021}\\
{}&              &                               & 05/11/2022& 121.0 & 4.98 & 1537+279 & & 9 & \\
{KISSR967}& 16 06 31.602 & 29 27 56.112 & 17/09/2022& 69.1 & 1.54 & 1603+301 & 0.60 & 7 & \citet{Kharb2021}\\
{KISSR1154}& 11 56 32.871 & 42 59 39.270 & 16/09/2022& 40.5 & 1.54 & 1147+438 & 1.26&  7 & \citet{Kharb2021}\\
{KISSR1219}& 12 09 08.809 & 44 00 11.500 & 06/11/2022& 69.4 &  1.54 & 1218+444 & 2.22 & 9 & \citet{Kharb2017a}\\
{KISSR1494}& 13 13 25.849 & 43 32 14.790 & 17/10/2022& 52.7 & 1.54 & 1325+436 & 2.53 & 8 & \citet{Kharb2015}\\
\enddata
\tablecomments{Columns 1, 2 and 3: source name and right ascension, declination of the sources. Column 4: date of observation in day, month and year. Column 5: Separation between the source and the Sun in degrees. Columns 6, 7, 8, and 9: the observing frequency in GHz, phase reference calibrator name, its separation from the KISSR source in degrees, and the number of VLBA antennas used in the observations, respectively. {Column~10: References for epoch-I observations.} ${\ast}$Epoch-II results from KISSR872 have been presented in \citet{Kharb2024}. }
\end{deluxetable*}

\section*{Epoch-I VLBI Observations}
The {first epoch of} phase-referenced VLBA observations revealed one-sided pc-scale jets in $63$\% or 5 out of the 8 detected sample sources \citep{Kharb2021}. 
One source, viz. KISSR1321, which had the lowest VLA FIRST flux density of 1.6~mJy, was not detected at either frequency with the VLBA. Double radio cores consistent with the presence of binary SMBHs were detected in only 1 source, viz. KISSR102 \citep{Kharb2020}. Explaining jet one-sidedness by Doppler boosting or dimming effects {in approaching or receding jets, respectively}, implied jet speeds in the range of $0.003c - 0.75c$ for assumed jet inclination angles of $\gtrsim50^\circ$; this limit was obtained from their type~2 classification {and the expected half opening angle of the dusty torus} \citep{Kharb2021}. Invoking free-free absorption for the missing counterjet emission presented a difficulty because of the low filling factors of the absorbing gas clouds and the large ($\sim100$~pc-scale) jets observed in several of the sources \citep[see][]{Kharb2019}.

Steep radio spectral indices were either observed in both the jets and the `cores' of several sources (e.g., KISSR102, KISSR434, KISSR872) or implied by the non-detection of cores and jets in the higher frequency (i.e., 5~GHz) observations (e.g., KISSR618, KISSR967, KISSR1154, KISSR1219, KISSR1494). In the literature, both flat or inverted radio `core' spectra \citep[e.g.,][]{Mundell2000, Ho2008, Falcke2000, Kharb2017b} and steep `core' spectra \citep[e.g.,][]{Roy00, Bontempi12, Falcke2000, Giroletti2009, Chiaraluce2019} have been observed in the VLBI images of Seyfert and LINER galaxies. 

In this paper, we present the results from second-epoch VLBI observations of the 8 (out of the original 9) sources that were detected in the first epoch \citep{Kharb2021}. Here, similar to our previous work, we adopt the cosmology with H$_0$ = 73~km~s$^{-1}$~Mpc$^{-1}$, $\Omega_{mat}$ = 0.27, $\Omega_{vac}$ =  0.73. Spectral index $\alpha$ is defined such that flux density at frequency $\nu$ is $S_\nu \propto \nu^{\alpha}$.

\section{VLBA Observations} \label{sec:data}
Phase-referenced VLBA observations of 8 KISSR galaxies were carried out at 1.5 GHz (Project ID: BK246). As only 3 of the original 9 sources, viz. KISSR102, KISSR434, and KISSR872 were detected at 5~GHz in the first epoch, we re-observed only these 3 sources at 5~GHz in the second epoch. The VLBA observations were carried out between September 1, 2022 and November 29, 2022. 
Between 7 and 9 VLBA antennas participated in these experiments. Data were recorded in both bands with a total bandwidth (BW) of 256~MHz (8 intermediate frequencies or IFs with a BW of 32 MHz each) and an aggregate bit rate of 2048 Mbps. Nearby compact sources with accurate small positional uncertainties were used as phase-referencing calibrators (see Table~\ref{tab:obs}). The bright calibrators 4C39.25 and 3C345 were used as fringe-finders. The targets and the phase reference calibrators were observed in a `nodding' mode in a 5 minute cycle (2 minutes on calibrator and 3 minutes on source), for good phase calibration. 

In Table~\ref{tab:obs}, we also note the separation between the sources and the recently active Sun. These separations typically ranged between 40$\degr$ and 70$\degr$. The separation was the smallest ($\sim$30$\degr$) for KISSR102. The Sun's proximity might have affected the phases in the visibility data of KISSR102 (this source shows the smallest dynamic range in these observations). The small separation of its phase reference calibrator (1.78$\degr$) however, and the application of ionospheric corrections during the initial calibration using the Astronomical Image Processing Software \citep[{\tt AIPS;}][]{Greisen2003} task {\tt TECOR}, were able to  mitigate these effects, as was evident from the detection of well correlated phases in KISSR102. The experiment’s minimum observing elevation was limited to 8 degrees, in accordance with the constraints applied by the SCHED software while preparing the schedule files.

The data reduction was carried out using the {\tt VLBARUN}\footnote{\url{https://www.aips.nrao.edu/cgi-bin/ZXHLP2.PL?VLBARUN}} procedure in {\tt AIPS}. {\tt VLBARUN} uses the VLBA calibration procedures ({\tt VLBAUTIL}) to calibrate the VLBA data. The calibration steps included a correction of the ionospheric dispersive delays using the task {\tt TECOR} and Earth Orientation Parameters (EOPs) corrections using the task {\tt CLCOR}. After this, amplitude and instrumental delay calibration (using the task {\tt FRING} on a scan of the fringe-finder), followed by a calibration of the bandpass shapes, antenna rates and phases, was carried out. After parallactic angle corrections, the data were fringe-fit with task {\tt FRING} for a solution interval of 2 minutes. This typically resulted in less than 1\% of failed solutions. The fringe-fitting combined the IFs to increase the signal to noise ratio of delay fits. The Stokes RR and LL signals were however, calibrated independently. The phase solutions were linearly interpolated in time between calibrator scans, with the caveat that this approach may be inappropriate if the phase varies non-linearly on timescales shorter than the phase-referencing cycle time, particularly due to the proximity of the Sun. Los Alamos (LA) was chosen as the reference antenna for calibrating the various datasets. Data were not flagged, and the phase reference calibrators, chosen as compact sources on VLBI scales, were not self-calibrated during {\tt VLBARUN}. 

Finally, the 64 spectral channels were averaged, and individual sources were `SPLIT' from the multi-source data after the application of the calibration tables. We note that these data-reduction steps, as well as the choice of calibrators, were identical to those used in Epoch-I observations.

\begin{figure*}
\centering
\includegraphics[width=8.5cm,trim=100 140 0 150]{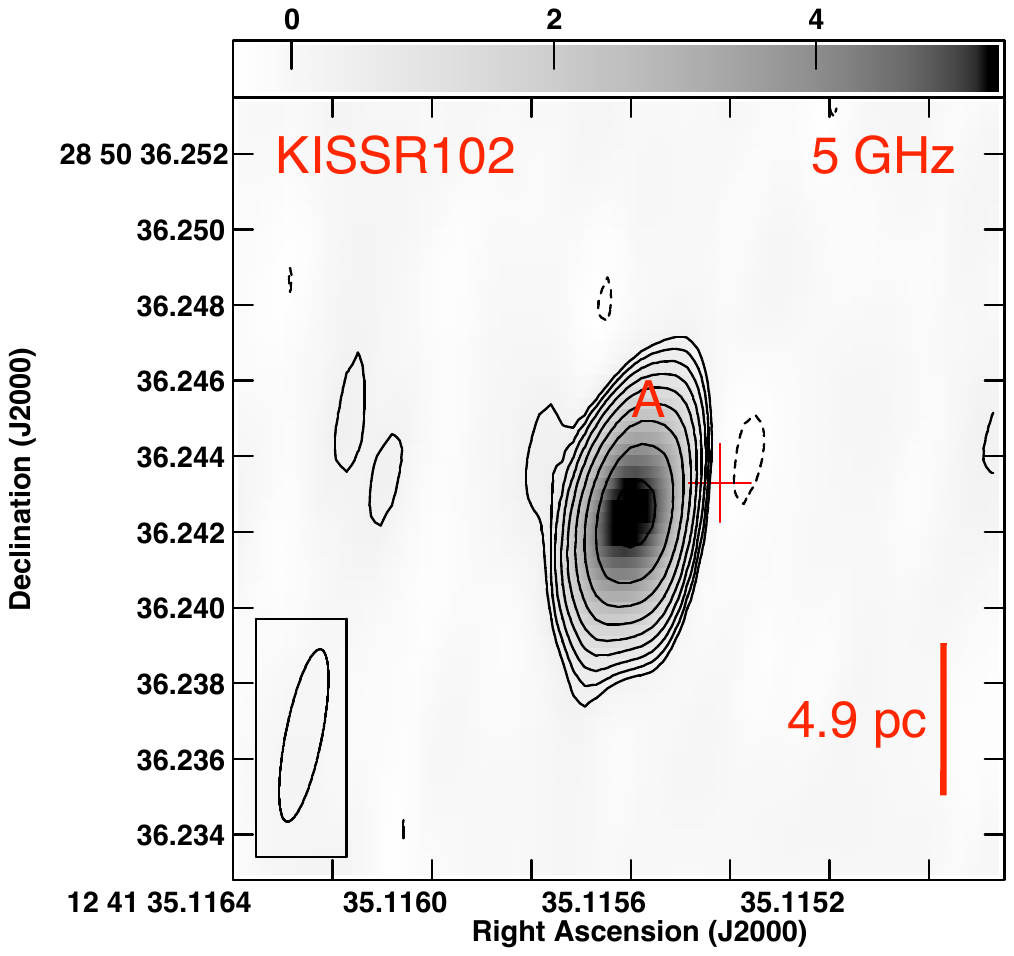}
\includegraphics[width=7.3cm,trim=70 140 100 150]{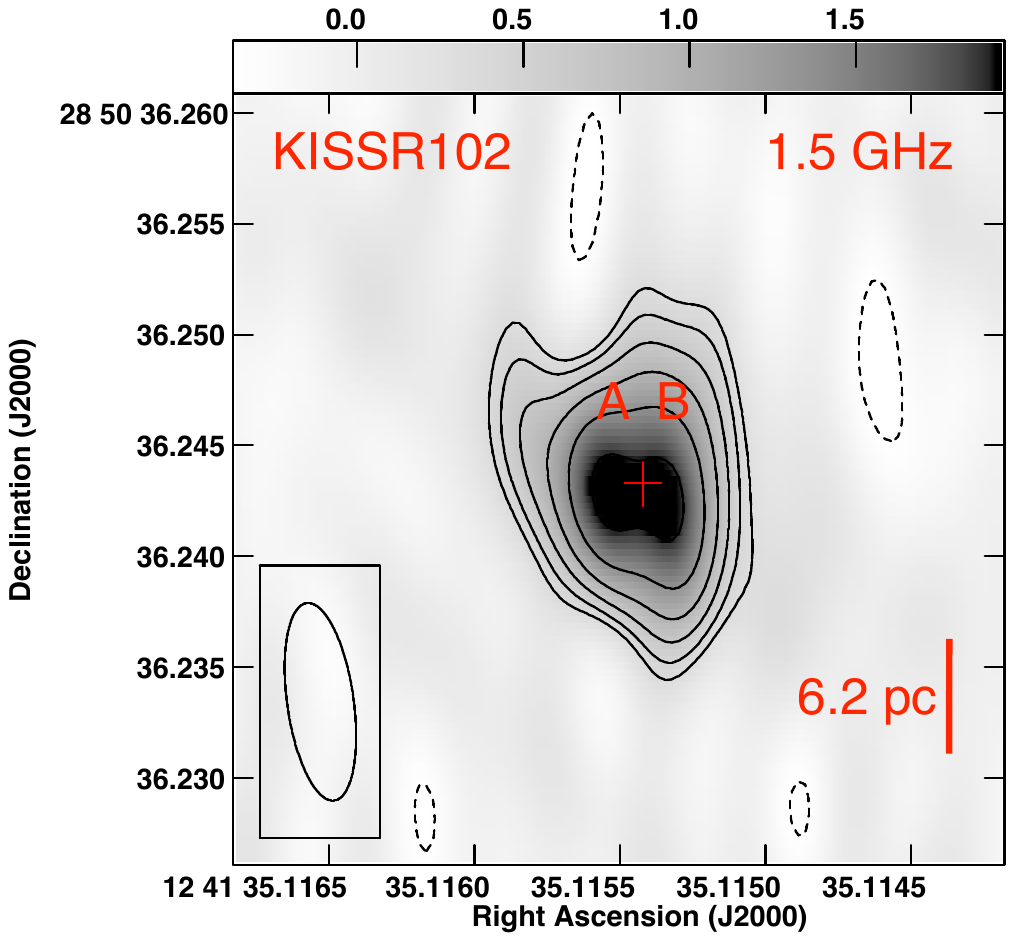}
\includegraphics[width=11cm,trim=0 200 0 200]{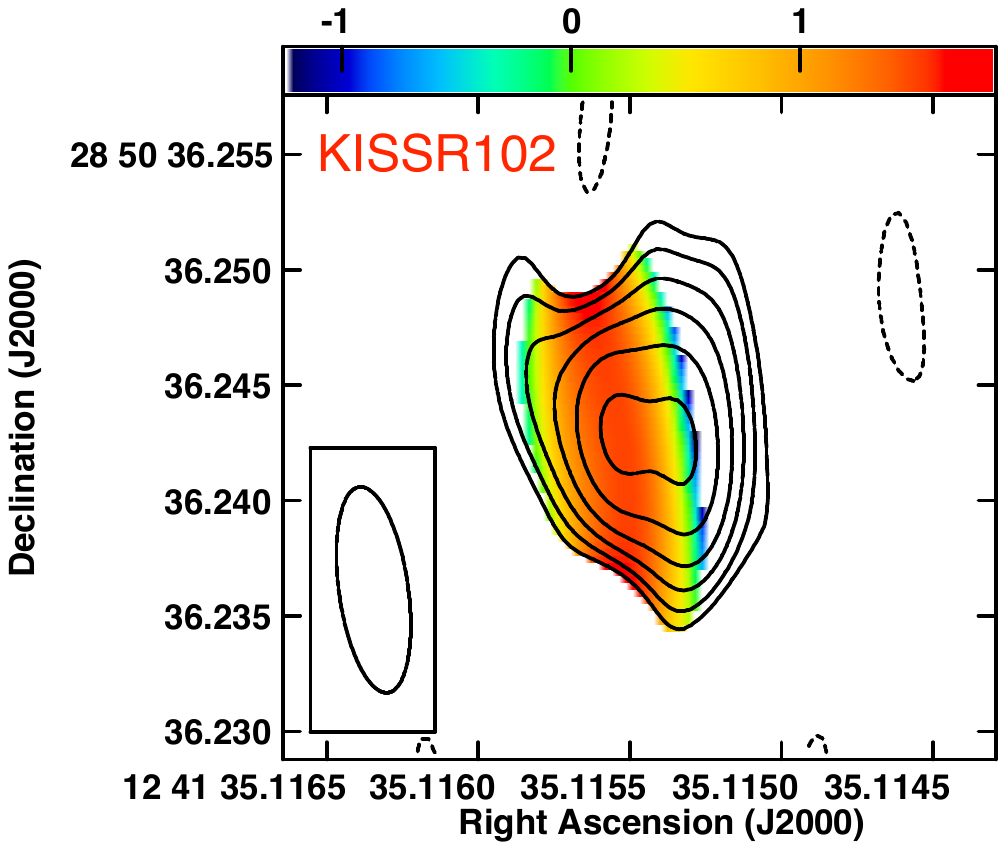}
\caption{\small (Top left) 5~GHz VLBA image of KISSR102. The gray-scale varies from $-0.36$ to 5.35~mJy~beam$^{-1}$. The contour levels are $59.17\times(\pm5.6, 11.3, 22.5, 45, 90)$~$\mu$Jy~beam$^{-1}$. The beam is of size $4.6\times0.96$~mas at PA$=-11.2\degr$. (Top right) 1.5~GHz VLBA image of KISSR102. The gray-scale varies from $-0.32$ to 1.82~mJy~beam$^{-1}$. The contour levels are $20.26\times(\pm16, 22.5, 32, 45, 64, 90)$~$\mu$Jy~beam$^{-1}$. The beam is of size $8.4\times2.8$~mas at PA$=4.4\degr$. In both the panels, the cross symbol denotes the Gaia position of the galaxy: RA 190.396314252$\degr$, Dec 28.843400915$\degr$ with respective errors of 1.03 mas and 0.82 mas. The size of the cross is 2 times the positional uncertainty noted here. (Bottom) $1.5-5$~GHz spectral index image in color superimposed by 1.5~GHz contours. The contour levels are $21.27\times(\pm16, 22.5, 32, 45, 64, 90)$~$\mu$Jy~beam$^{-1}$. The beam is of size $9\times3$~mas at PA$=8\degr$. }
\label{fig:k102}
\end{figure*}

\begin{figure*}
\centering
\includegraphics[width=10cm]{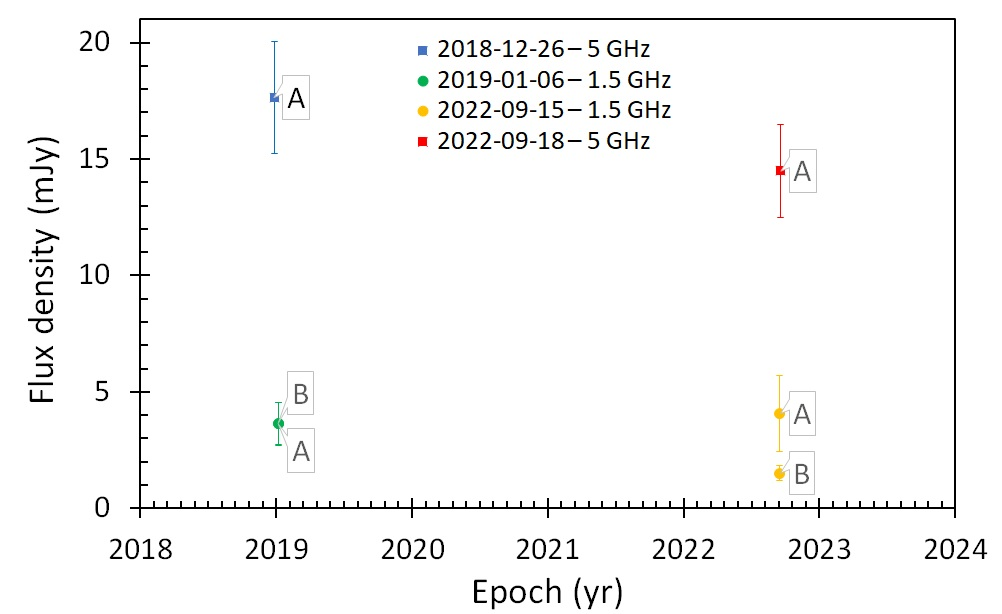}
\caption{\small The flux density of the components detected in the radio images of KISSR102 at 1.54 GHz (green and orange circles) and 4.98 GHz (blue and red squares). Component A is the core region of KISSR102, and B is the jet component.}
\label{fig:KISSR102_FluxDens}
\end{figure*}

\begin{figure*}
\centering
\includegraphics[width=10cm]{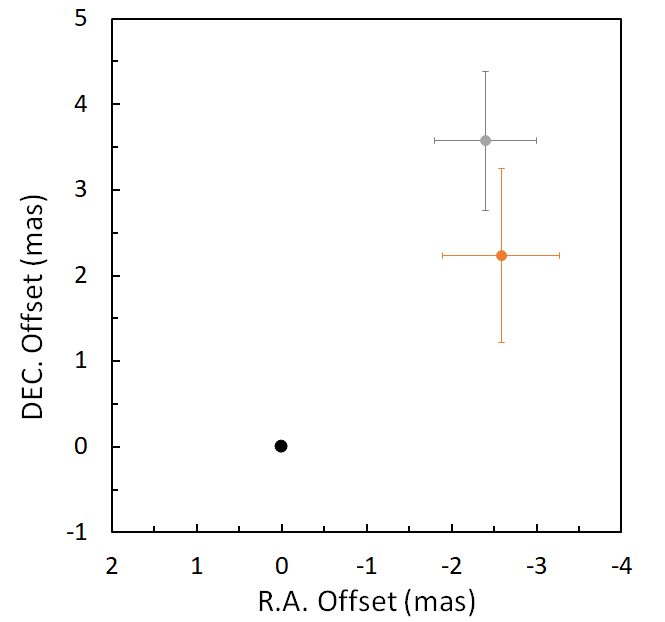}
\includegraphics[width=10cm]{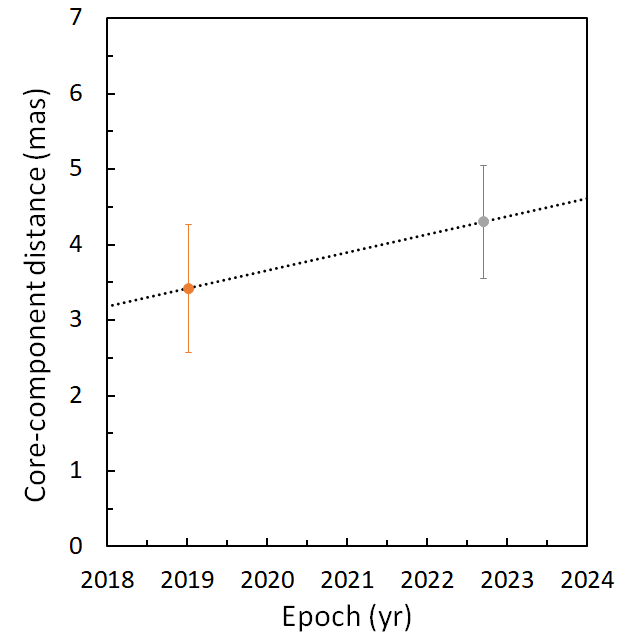}
\caption{\small (Top panel) Offsets in right ascension and declination of the jet component B in KISSR102 in 2019 and 2022 (orange and gray points, respectively) in relation to the core (component A), marked by the black circle. (Bottom panel) The angular distance between A and B in the same two epochs. The dotted line refers to a proper motion of 0.24 mas yr$^{-1}$ (apparent speed of 1.05$c$).}
\label{fig:KISSR102_RADEC_rxt}
\end{figure*}

\begin{figure*}
\centering
\includegraphics[width=10cm,trim=150 240 150 240]{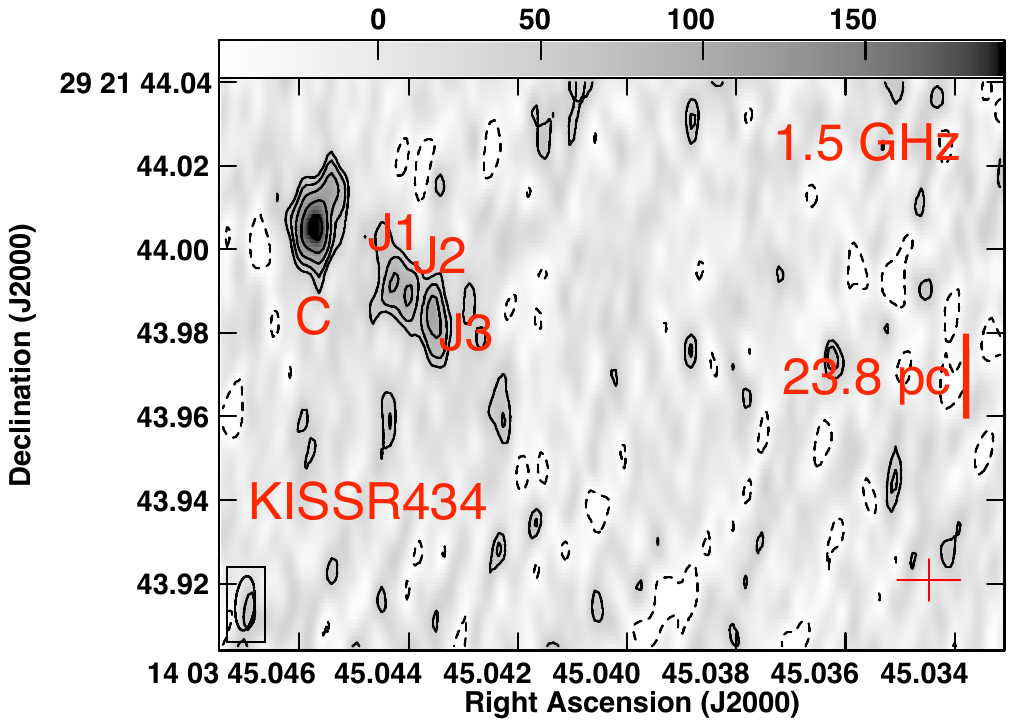}
\caption{\small 1.5~GHz VLBA image of KISSR434. The gray-scale varies from $-47.7$ to 191.5~$\mu$Jy~beam$^{-1}$. The contour levels are $2.12\times(\pm22.5, 32, 45, 64, 90)$~$\mu$Jy~beam$^{-1}$. The beam is of size $13.03\times4.57$~mas at PA$=-4.9\degr$. The cross symbol denotes the Gaia position of the galaxy: RA 210.937643641$\degr$, Dec 29.362200227$\degr$ with respective errors of 1.96 mas and 2.99 mas. The size of the cross symbol is 5 times this positional uncertainty.}
\label{fig:k434}
\end{figure*}

\begin{figure*}
\centering
\includegraphics[width=16cm,trim=50 160 50 160]{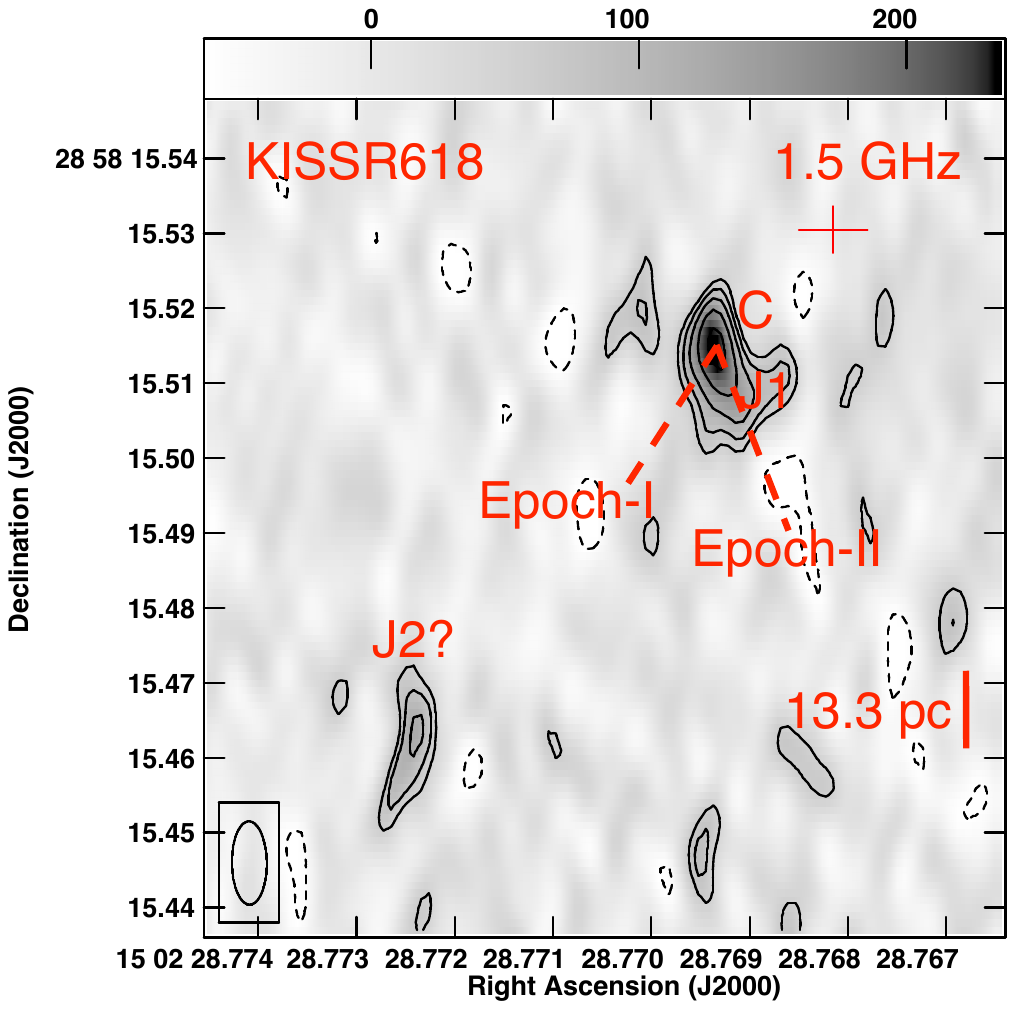}
\caption{\small 1.5~GHz VLBA image of KISSR618. The gray-scale varies from $-59.6$ to 234.2~$\mu$Jy~beam$^{-1}$. The contour levels are $2.59\times(\pm22.5, 32, 45, 64, 90)$~$\mu$Jy~beam$^{-1}$. The beam is of size $11.11\times4.65$~mas at PA$=0.6\degr$. The cross symbol denotes the Gaia position of the galaxy: RA 225.619867285$\degr$, Dec 28.970980695$\degr$ with respective errors of 1.23 mas and 1.82 mas. The size of the cross symbol is 5 times this positional uncertainty.}
\label{fig:k618}
\end{figure*}

\begin{figure*}
\centering
\includegraphics[width=14cm,trim=50 190 50 190]{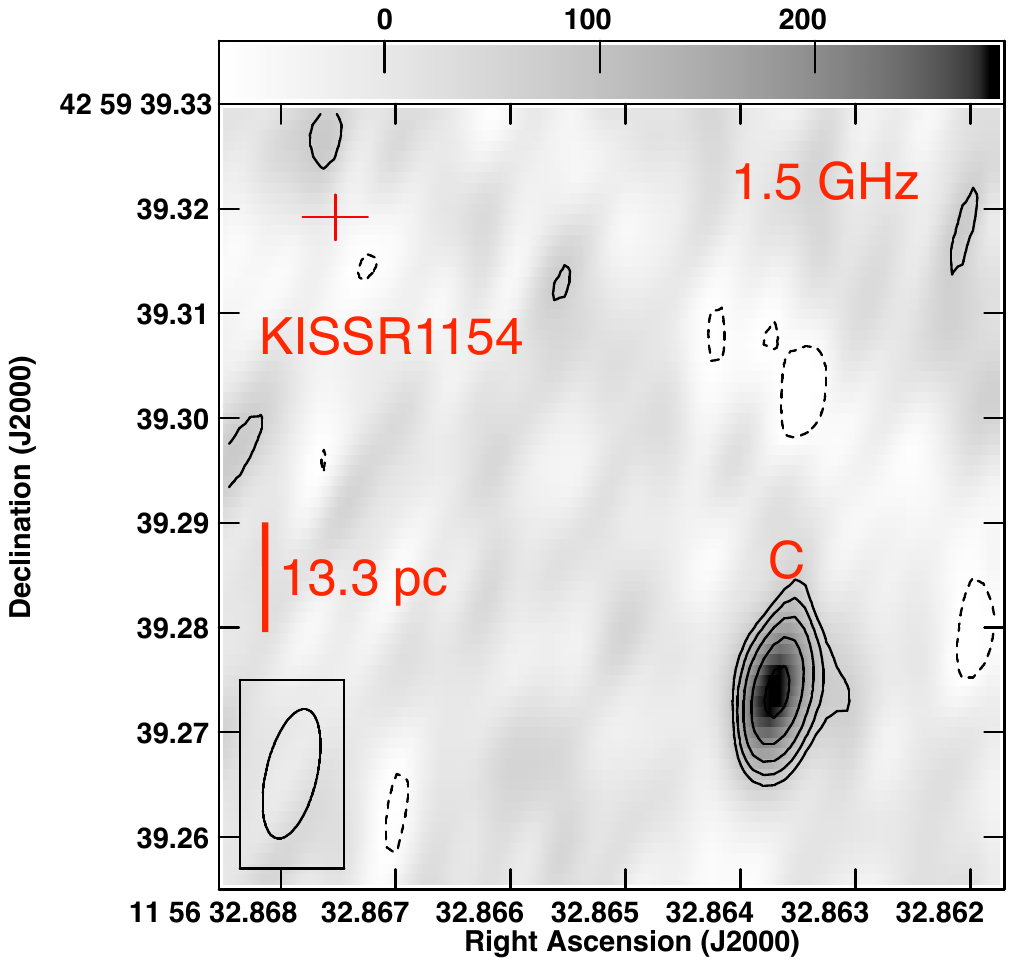}
\caption{\small 1.5~GHz VLBA image of KISSR1154. The gray-scale varies from $-70.4$ to 281.7~$\mu$Jy~beam$^{-1}$. The contour levels are $3.13\times(\pm22.5, 32, 45, 64, 90)$~$\mu$Jy~beam$^{-1}$. The beam is of size $12.62\times4.87$~mas at PA$=-13.1\degr$. The cross symbol denotes the Gaia position of the galaxy: RA 179.136948020$\degr$, Dec 42.994255321$\degr$ with respective errors of 4.36 mas and 6.19 mas. The size of the cross symbol represents this positional uncertainty.}
\label{fig:k1154}
\end{figure*}

\begin{figure*}
\centering
\includegraphics[width=14cm,trim=50 180 50 180]{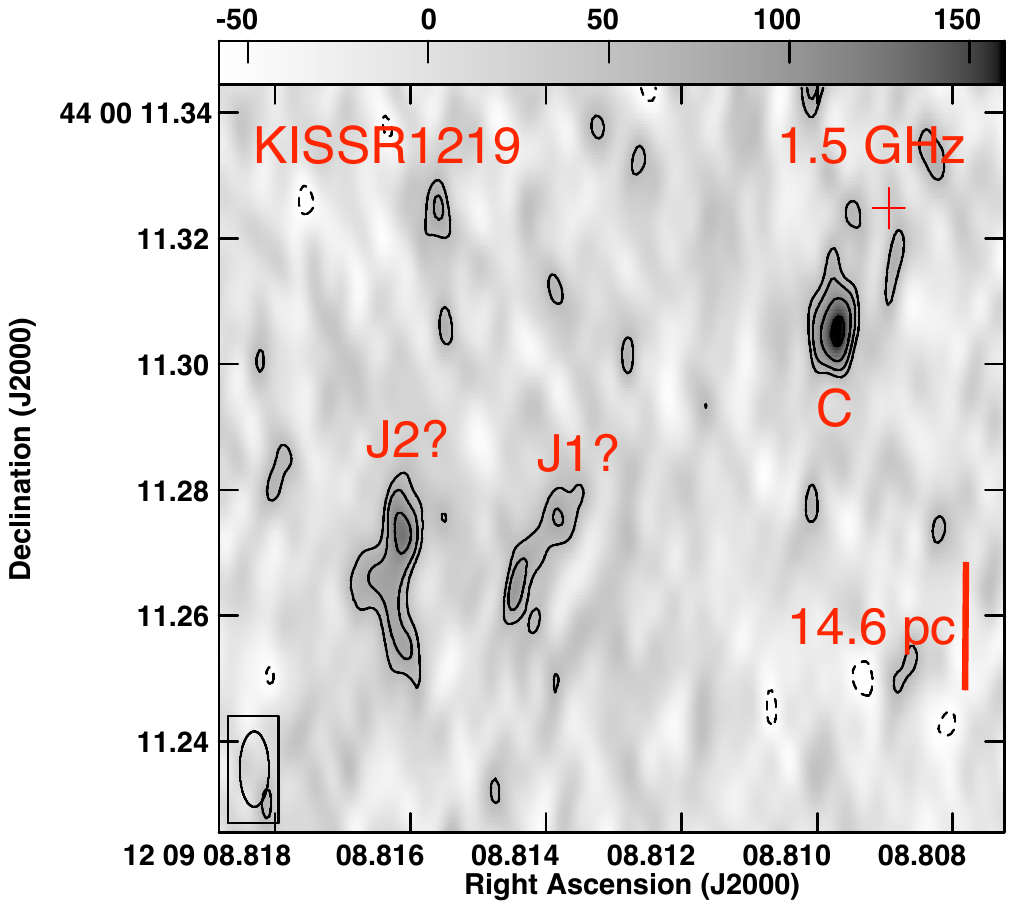}
\caption{\small 1.5~GHz VLBA image of KISSR1219. The gray-scale varies from $-56.0$ to 157.5~$\mu$Jy~beam$^{-1}$. The contour levels are $1.75\times(\pm32, 45, 64, 90)$~$\mu$Jy~beam$^{-1}$. The beam is of size $11.98\times4.65$~mas at PA$=0.1\degr$. The cross symbol denotes the Gaia position of the galaxy: RA 182.286703941$\degr$, Dec 44.003145789$\degr$ with respective errors of 0.64 mas and 0.50 mas. The size of the cross symbol is 10 times this positional uncertainty.}
\label{fig:k1219}
\end{figure*}

Imaging was carried out using the {\tt AIPS} task {\tt IMAGR}. Natural weighting with a {\tt ROBUST} parameter of +5 was used for the imaging (Figures~\ref{fig:k102}-\ref{fig:k872}). {Due to the radio faintness of the sources, no self-calibration was carried out. The peak intensity and total flux density with errors for compact components like the `core' were estimated using the Gaussian-fitting task {\tt JMFIT} in {\tt AIPS} (see Table~\ref{tab:results}). The final r.m.s. noise in the images was obtained using the {\tt AIPS} verbs {\tt TVWIN} and {\tt IMSTAT}. We obtained the total flux density of the extended features (e.g., `C+J', and J1, J2 in Table~\ref{tab:results}) using the {\tt AIPS} verb {\tt TVSTAT}. The error in the total flux density was typically $\le5$\%.}

We created the $1.5–5$~GHz spectral index image for KISSR102 by first making differently weighted images using {\tt ROBUST} $=-5$ (pure uniform weighting) at 1.5 GHz and {\tt ROBUST} $=+5$ (natural weighting) at 5~GHz and then convolving  both images with a synthesized beam of 9 mas $\times$ 3 mas at PA $=8\degr$. The {\tt AIPS} task {\tt COMB} was used to create the spectral index image (Figure~\ref{fig:k102}). Pixels with flux density values below 3 times the r.m.s. noise were blanked using {\tt COMB}. A spectral index noise image was created as well.

\section{VLA Observations}
Observations were also carried out for 4 KISSR sources, {viz. KISSR434, KISSR618, KISSR872, and KISSR967} on February 23, 2021 at 1.5~GHz with the Karl G. Jansky Very Large Array (VLA) in the A-array configuration, with a bandwidth of 1~GHz. The remaining 4 sources could not be observed due to the low scheduling priority of the project; data from these sources will be acquired in the near future. For the VLA experiment, 3C286 was used as the flux density and polarization angle calibrator while the unpolarized source J1407+2827 was used as the polarization leakage calibrator. J1407+2827 was used as the phase calibrator for KISSR434, J1513+2338 for KISSR618, and J1609+2641 for KISSR872 and KISSR967, respectively.  

Each source was observed for about $\sim15$ minutes. The initial calibration on the data was performed by using the CASA-based automated VLA pipeline which includes flagging, and initial bandpass and gain calibrations. We performed additional flagging, final gain calibration, and polarization calibration following standard procedures. The calibrated visibility data for the sources were then {\tt SPLIT} and fast Fourier transformed using the {\tt tclean} task in CASA with the Multi-term Multi-frequency Synthesis ({\tt MTMFS}) deconvolution algorithm, and a Briggs' weighting with {\tt ROBUST} parameter of $+0.5$ ({\tt ROBUST} 0 for KISSR434). We performed self-calibration on KISSR434 and KISSR967 with 3 rounds of phase-only and 2 rounds of amplitude+phase calibration. For KISSR618 and KISSR872, self-calibration could not be carried out due to too many failed solutions because of the faintness of the sources. The resultant Stokes I, Q, U images were used to make the total intensity, linear polarization and fractional polarization images (see Figures~\ref{fig:fig10}-\ref{fig:fig12}).

As the {\tt MTMFS} algorithm was used with {\tt nterms=2}, we also obtained the spectral index and error images. The task {\tt COMB} was used for creating the polarization images by blanking pixels with less than 3$\sigma$ signal in Stokes Q and U images, and errors greater than 10$\degr$ in the polarization angle image. The fractional polarization was created after blanking pixels with greater than 20\% error in fractional polarization. The spectral index images presented in Figure~\ref{fig:fig10} were created after using spectral index noise images and blanking errors greater than 0.5, 0.2, 0.5, and 0.3 for KISSR434, KISSR872, KISSR618 and KISSR967, respectively. {Pixels with flux density values below 3 times the r.m.s. noise in the total intensity image were also blanked using the task {\tt COMB}.}

\begin{deluxetable*}{lccccccccc}
\tabletypesize{\scriptsize}
\tablewidth{0pt} 
\tablecaption{Results from the VLBA observations\label{tab:results}}
\tablehead{\colhead{Source name}&\colhead{$z$}&\colhead{Type}&\colhead{Component}&\colhead{$\mathrm{I_{peak}}$} & \colhead{$\mathrm{S_{total}}$} & \colhead{r.m.s.} & \colhead{Scale} & \colhead{Extent} & \colhead{Extent}\\
\colhead{\& Freq band} & \colhead{} & \colhead{} & \colhead{} & \colhead{$\mu$Jy~beam$^{-1}$} & \colhead{$\mu$Jy}& \colhead{$\mu$Jy~beam$^{-1}$}& \colhead{pc/mas} & \colhead{mas} & \colhead{parsec}}
\colnumbers
\startdata
{KISSR102 L-band}& 0.066320 & LINER & A & $1924\pm98$ & $2538\pm204$ & 100.6   & 1.237 & {4.30} & {5.40} \\
& &  & B & $1697\pm97$ & $2408\pm214$ & \nodata & \nodata & \nodata&\nodata \\
{KISSR102 C-band}&          &       & A & $6250\pm64$ & $13880\pm197$ & 75.65 & \nodata&\nodata &\nodata \\
{KISSR434 L-band}& 0.064128 & Sy 2 & C & $208\pm26$ & $475\pm80$ & 24.00 & 1.189 & 66.48 & 79.04 \\
        & & & C+J & \nodata & 497 &\nodata & \nodata & \nodata & \nodata\\
        & & & J1+J2+J3 & \nodata & 573 &\nodata & \nodata & \nodata & \nodata \\
{KISSR434 C-band}&  &  & N & \nodata & \nodata & 21.14 & \nodata & \nodata & \nodata \\
{KISSR618 L-band}& 0.072940 & Sy 2 & C & $254\pm25$ & $435\pm63$ & 25.13 & 1.332 & 77.95 & 103.83 \\
          & & & C+J1 & \nodata & 510&\nodata & \nodata & \nodata & \nodata\\
          & & & J2 & \nodata & 200 &\nodata & \nodata & \nodata & \nodata\\
{KISSR872$^{\ast}$ L-band}& 0.083064 & LINER & C+J1 & $475\pm27$ & $1601\pm116$ & 29.54 & 1.497 & \nodata&\nodata \\
{KISSR872$^{\ast}$ C-band}& \nodata & \nodata & C & $334\pm22$ & $472\pm49$ & 18.50 & \nodata & \nodata&\nodata \\
{KISSR967 L-band}& 0.092067 & LINER & N &\nodata&\nodata&36.98 & 1.648 &\nodata &\nodata \\
{KISSR1154 L-band}& 0.072032 & Sy 2 & C & $313\pm33$ & $385\pm66$ & 31.81 & 1.329 & \nodata & \nodata\\
           & & & C+J & \nodata & 387 &\nodata & \nodata & \nodata & \nodata \\
{KISSR1219 L-band}& 0.037580 & Sy 2 & C & $168\pm24$ & $252\pm56$ & 23.89 & 0.729 & 89.98 & 65.60\\
           & & & C+J & \nodata &267 &\nodata & \nodata & \nodata & \nodata\\
           & & & J1  & \nodata &235 &\nodata & \nodata & \nodata & \nodata\\
           & & & J2 & \nodata &389 &\nodata & \nodata& \nodata & \nodata\\
{KISSR1494 L-band}& 0.057446 & Sy 2 & C & $321\pm27$ & $691\pm81$ & 29.92 & 1.082 & 60.82 & 65.81\\
        & & & C+J1 & \nodata & 837 &\nodata & \nodata& \nodata & \nodata\\
        & & & J2   & \nodata & 180 &\nodata & \nodata& \nodata & \nodata 
\enddata
\tablecomments{{${\ast}$ Results for KISSR872 have been presented in \citet{Kharb2024}; in this paper, we reproduce the L-band images of KISSR872 for both epochs to highlight the changes in its jet base. Column 1: source name and observing frequency, L-band at 1.54~GHz, C-band at 4.98~GHz. Column 2: redshift. Column 3: AGN classification. Column 4: core (C), core+jet (C+J), individual jet components (J1, J2), N = no detection. `C+J' or `C+J1' denote the `core' along with unresolved jet emission. For KISSR102, components A and B as shown in Figure~\ref{fig:k102}. Columns 5 and 6: the peak intensity and total flux density. For the candidate cores, these were estimated using the {\tt AIPS} Gaussian-fitting task {\tt JMFIT}. The total flux density for C+J and other extended jet features was estimated using the {\tt AIPS} verb {\tt TVSTAT}, with the error typically being $\le5$\%.} Column 7: the final r.m.s. noise in the image. Column 8: spatial scale in parsec corresponding to 1 milli-arcsecond angular scale at the source distance. Columns 9 and 10: total extent of the radio (jet) emission in milli-arcseconds and parsecs, respectively.}
\end{deluxetable*}

\section{Results} \label{sec:results}
\begin{figure*}
\centering
\includegraphics[width=15cm,trim=50 180 50 180]{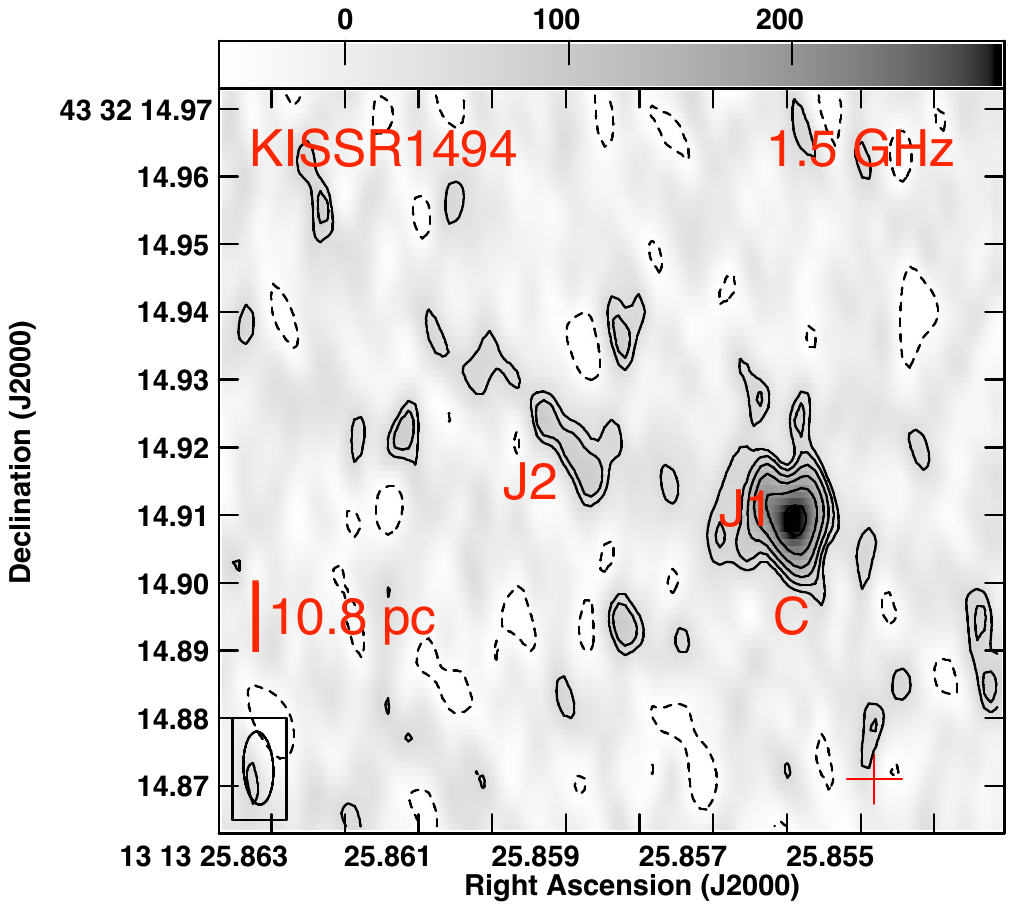}
\caption{\small 1.5~GHz VLBA image of KISSR1494. The gray-scale varies from $-53.3$ to 291.8~$\mu$Jy~beam$^{-1}$. The contour levels are $3.22\times(\pm16, 22.5, 32, 45, 64, 90)$~$\mu$Jy~beam$^{-1}$. The beam is of size $10.75\times4.40$~mas at PA$=2.8\degr$. The cross symbol denotes the Gaia position of the galaxy: RA 198.357728378$\degr$, Dec 43.537464172$\degr$ with respective errors of 0.36 mas and 0.40 mas. The size of the cross symbol is 10 times this positional uncertainty.}
\label{fig:k1494}
\end{figure*}

\begin{figure*}
\centering
\includegraphics[width=8.9cm,trim=60 220 50 200]{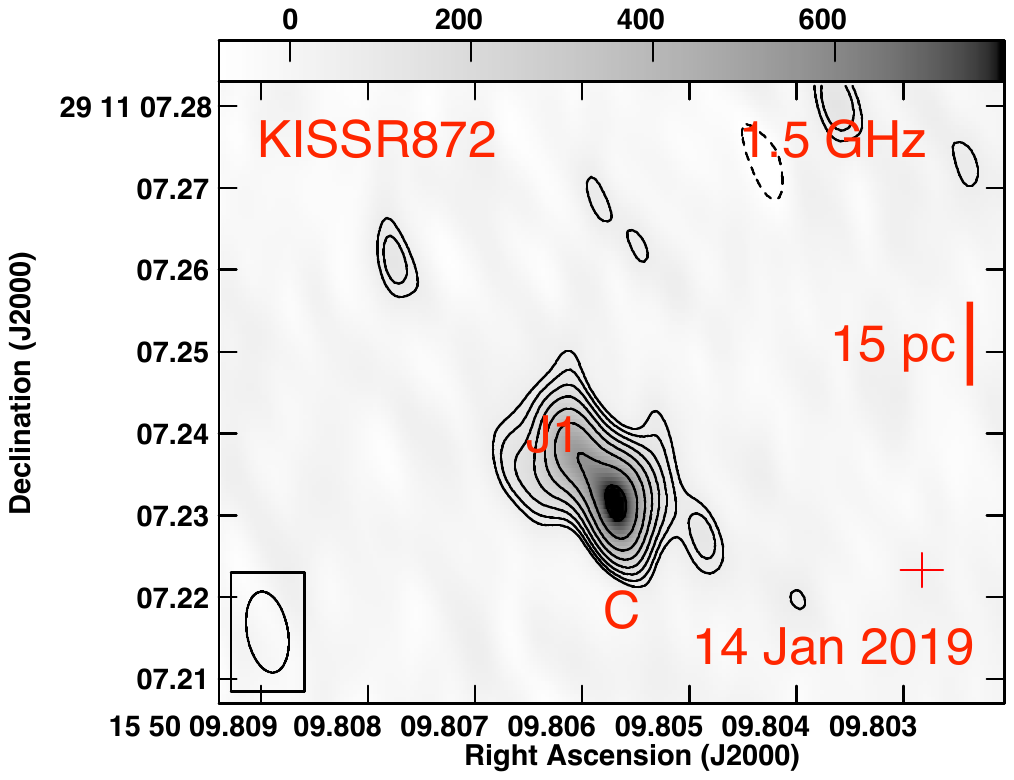}
\includegraphics[width=8.9cm,trim=50 220 60 200]{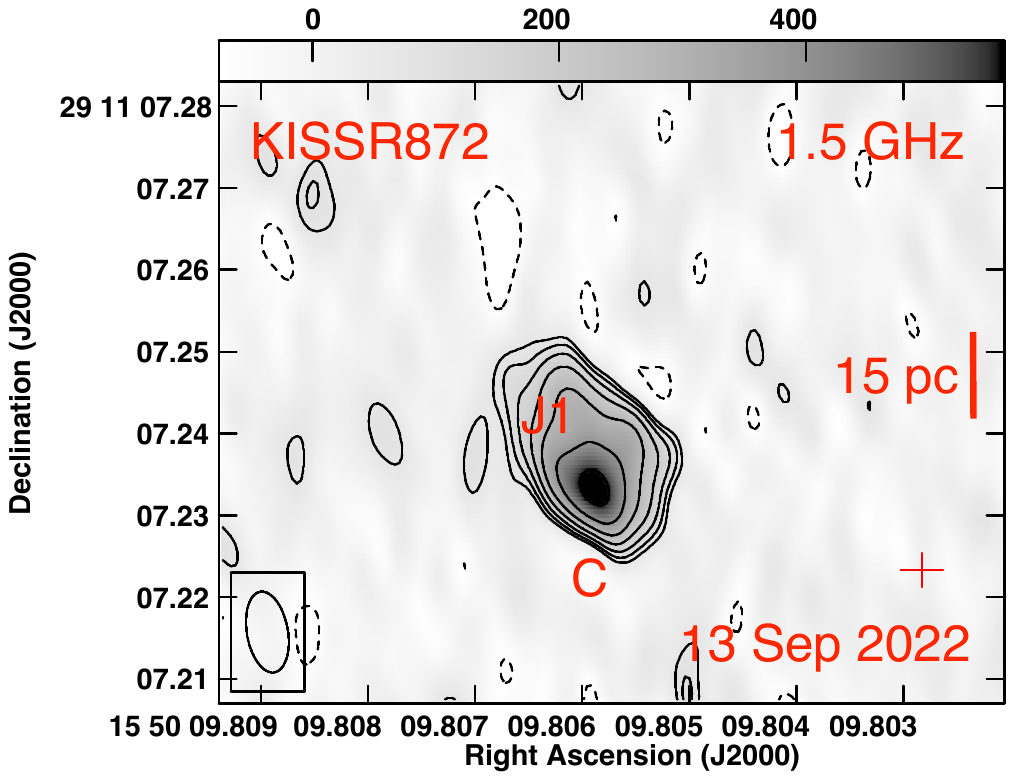} 
\caption{\small 1.5~GHz VLBA image of KISSR872 (left) from Epoch-I, 14 January 2019, and (right) from Epoch-II, 13 September 2022. The gray-scale varies from (left) $-73.3$ to 781.7~$\mu$Jy~beam$^{-1}$, and (right) $-72.8$ to 558.0~$\mu$Jy~beam$^{-1}$. The contour levels are (left) $8.64\times(\pm8, 11.3, 16, 22.5, 32, 45, 64, 90)$~$\mu$Jy~beam$^{-1}$, and (right) $6.17\times(\pm11.3, 16, 22.5, 32, 45, 64, 90)$~$\mu$Jy~beam$^{-1}$. The beam is of size $10\times5$~mas at PA$=10\degr$ in both the panels. The cross symbol denotes the Gaia position of the galaxy: RA 237.540845128$\degr$, Dec 29.185339806$\degr$ with respective errors of 0.41 mas and 0.51 mas. The size of the cross symbol is 10 times this positional uncertainty.}
\label{fig:k872}
\end{figure*}

\begin{figure*}
\centering
\includegraphics[width=8cm,trim=0 0 0 0]{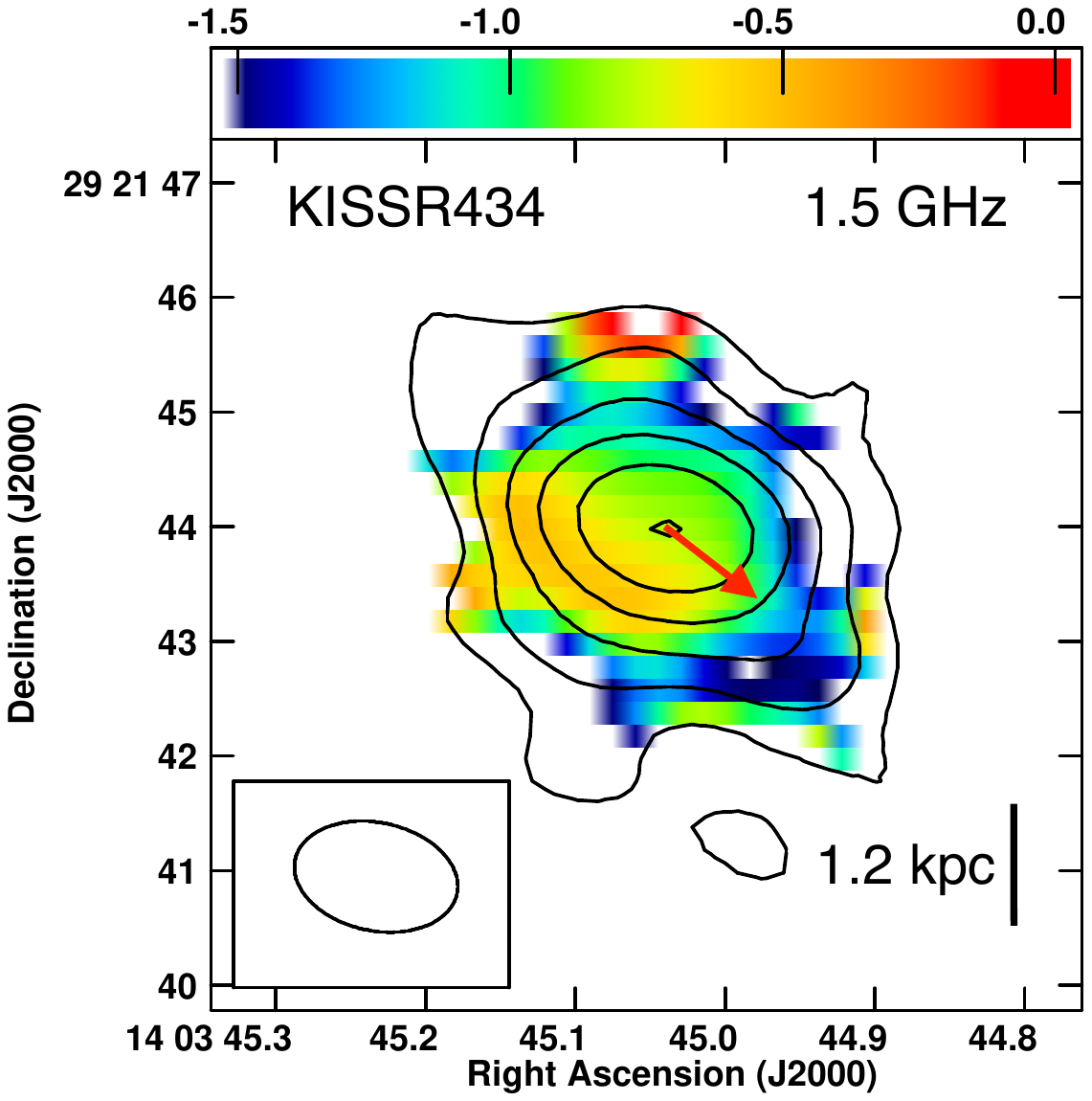}
\includegraphics[width=9.2cm,trim=0 55 0 0]{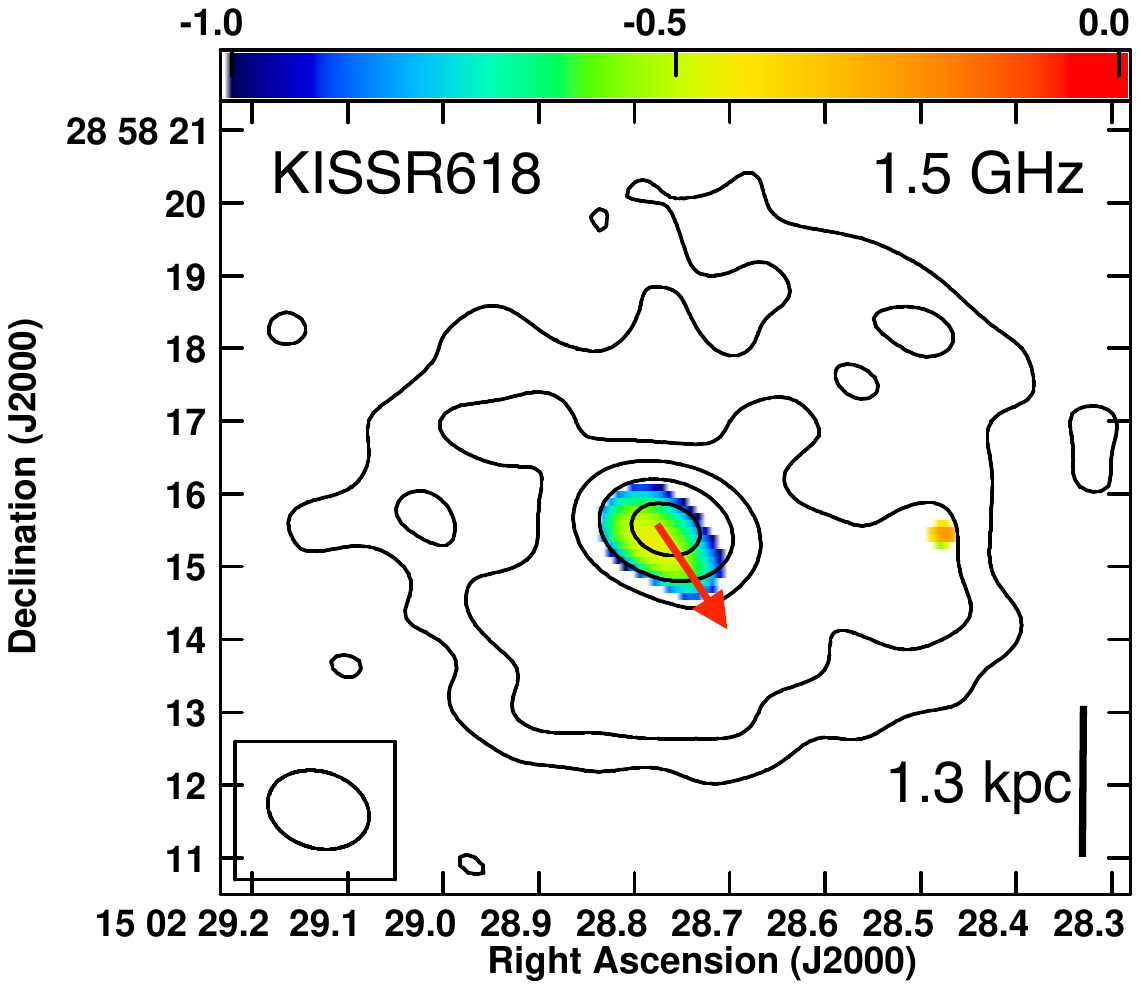}
\includegraphics[width=7.7cm,trim=50 110 0 210]{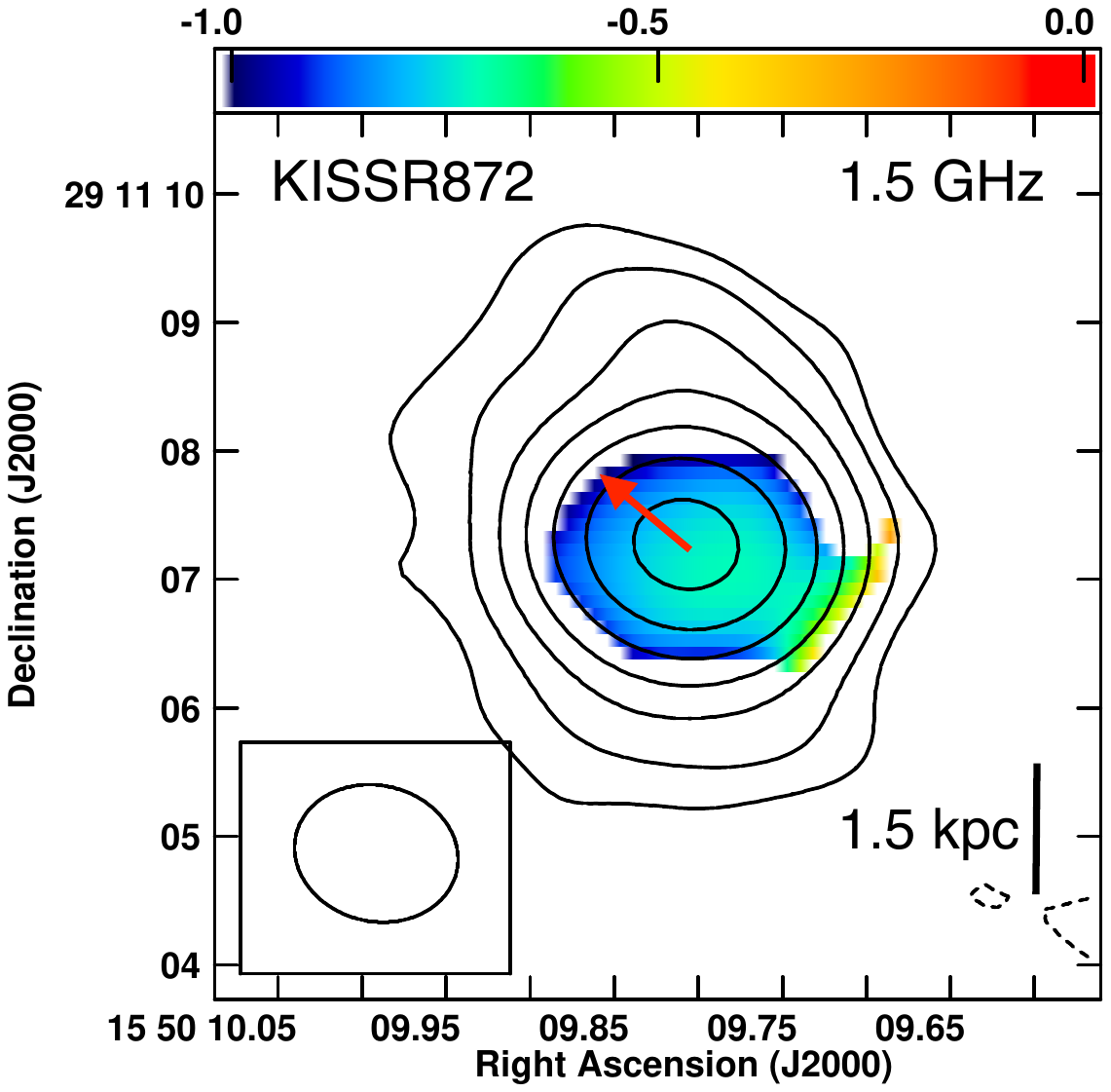}
\includegraphics[width=8.4cm,trim=10 160 40 140]{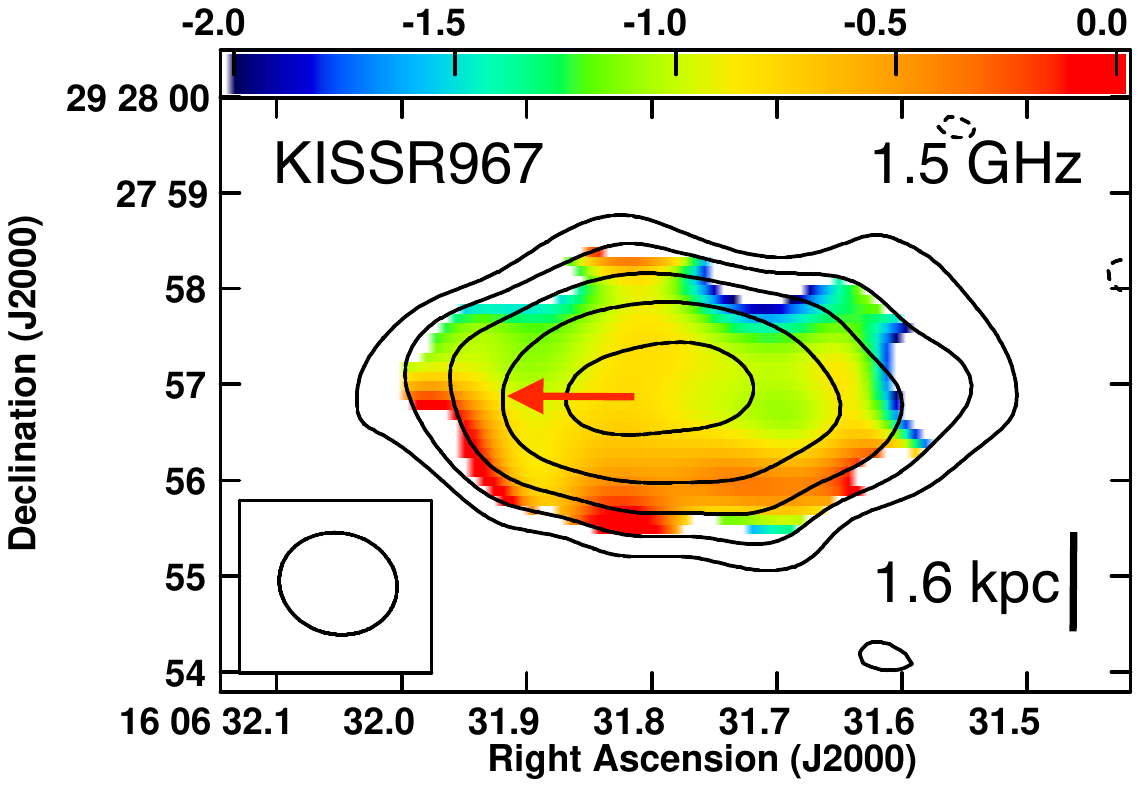}
\caption{\small VLA in-band (centred at 1.5~GHz) spectral index images in color of (top left) KISSR434, (top right) KISSR618, (bottom left) KISSR872, and (bottom right) KISSR967. The 1.5 GHz contours are $75\times(1, 2, 4, 8, 16, 32)$~$\mu$Jy~beam$^{-1}$ for KISSR434, $45\times(\pm1, 2, 4, 8, 16)$~$\mu$Jy~beam$^{-1}$ for KISSR618, $54\times(\pm1, 2, 4, 8, 16, 32, 64)$~$\mu$Jy~beam$^{-1}$ for KISSR872, and $45\times(\pm1, 1, 2, 4, 8, 16, 20)$~$\mu$Jy~beam$^{-1}$ for KISSR967. The red arrow denotes the position angle of the VLBI jet.}
\label{fig:fig10}
\end{figure*}

\begin{figure*}
\centering
\includegraphics[width=7.2cm,trim=80 0 20 80]{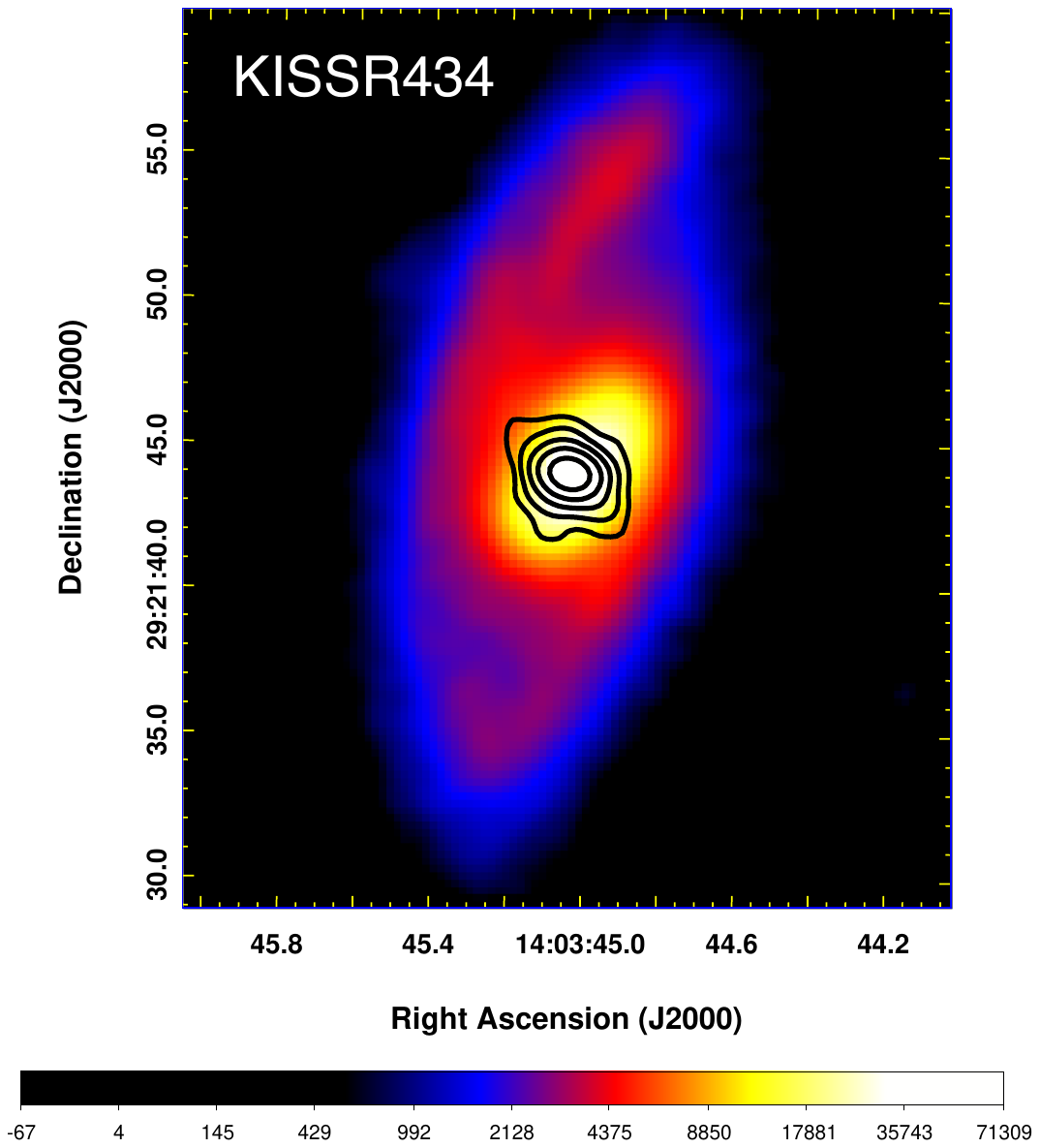}
\includegraphics[width=8cm,trim=20 90 100 100]{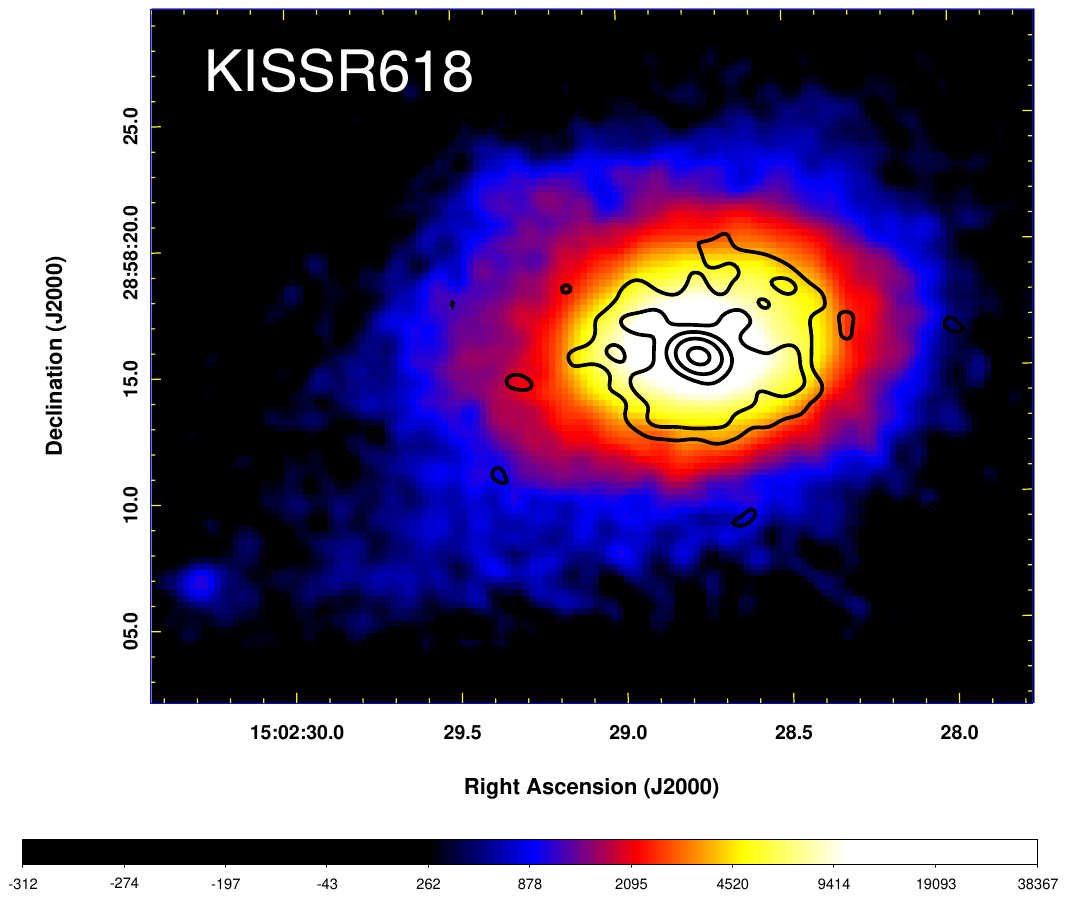}
\includegraphics[width=8cm,trim=30 110 0 150]{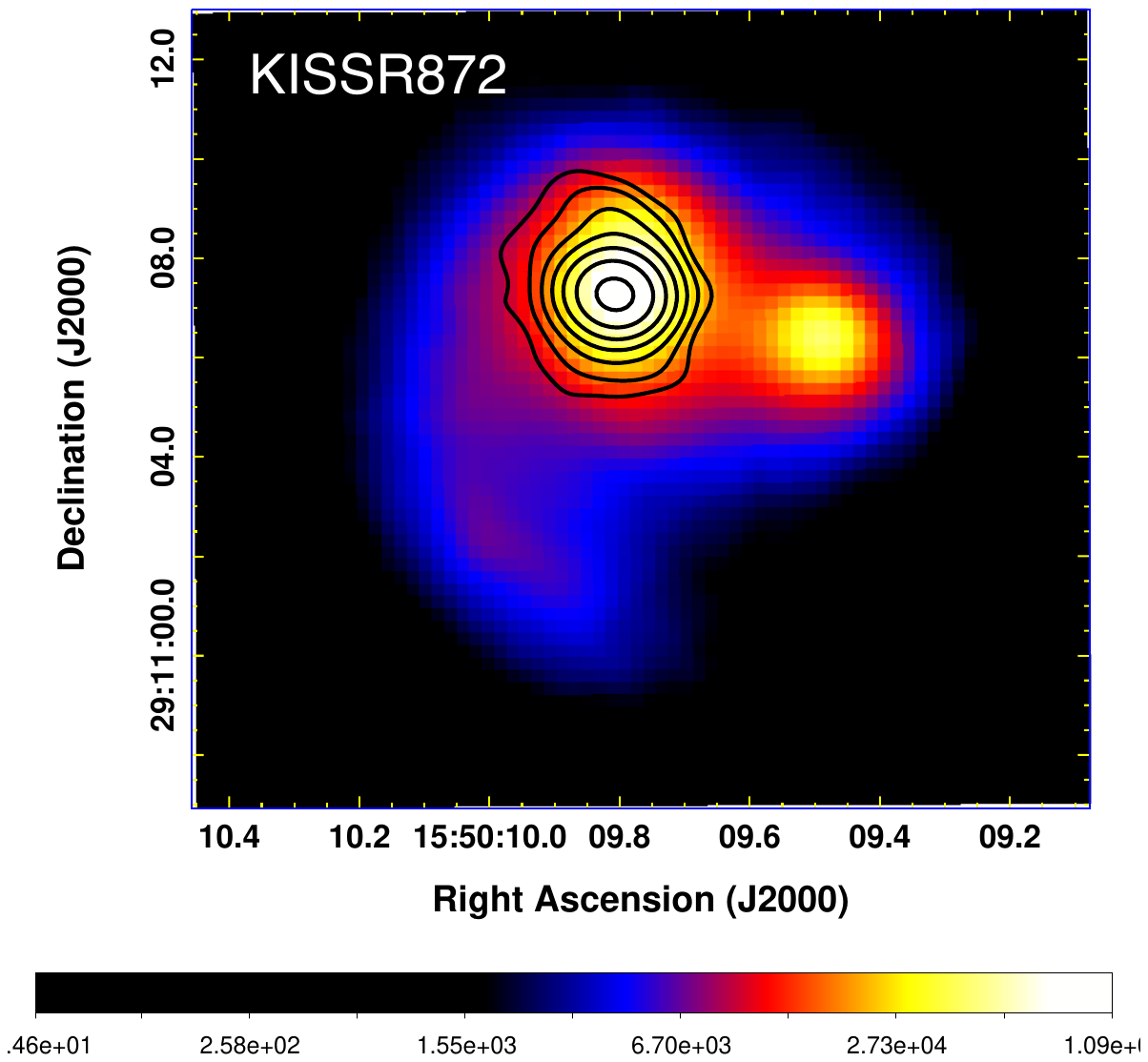}
\includegraphics[width=8.2cm,trim=0 160 60 210]{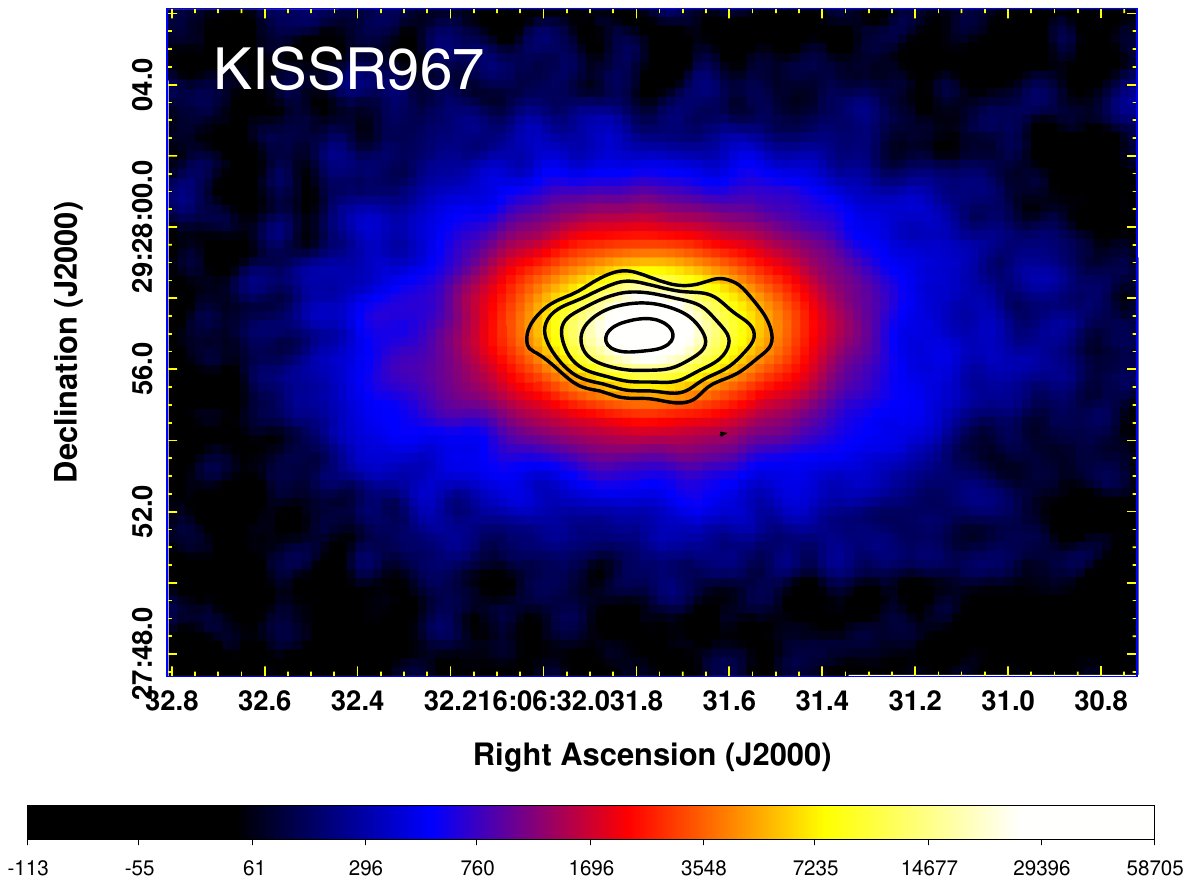}
\caption{\small {VLA radio contours at 1.5~GHz in black superimposed on} color images from Pan-STARRS i-band for KISSR434 (top left), KISSR618 (top right), KISSR872 (bottom left) and KISSR967 (bottom right). The contour levels are the same as noted in Figure~\ref{fig:fig10}. The FWHM of the synthesized beam is $1.44\times0.94$~arcsec at PA = 78.2$\degr$ for KISSR434, $1.42\times1.05$~arcsec at PA = 73.4$\degr$ for KISSR618, $1.28\times1.06$~arcsec at PA = 77$\degr$ for KISSR872, and $1.24\times1.06$~arcsec at PA = 77.9$\degr$ for KISSR967.}
\label{fig:fig11}
\end{figure*}

\begin{figure*}
\centering
\includegraphics[width=12cm,trim=100 200 100 140]{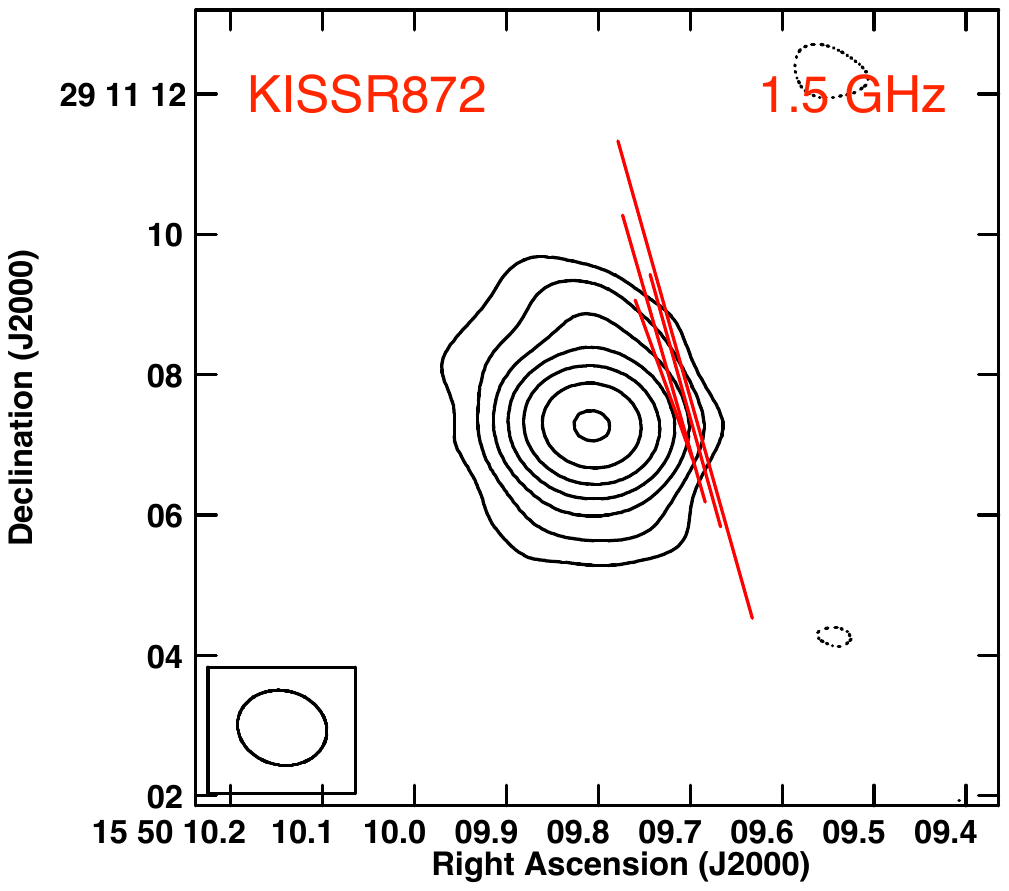}
\caption{\small {VLA 1.5~GHz image of KISSR872. The radio contours in black are overlaid with polarization vectors in red whose length is proportional to fractional polarization.} The contours are $45.36\times(\pm1.4,2.8,5.6,11.3,22.5,45,90)$~$\mu$Jy~beam$^{-1}$. 1 arcsec of the polarization vector corresponds to a fractional polarization of 6.25\%. The FWHM of the synthesized beam is $1.28\times1.06$~arcsec at PA = 77.1$\degr$.}
\label{fig:fig12}
\end{figure*}

The VLBA detected pc-scale emission at 1.5~GHz in 7 out of the 8 KISSR sources in these second epoch observations {(see Figures~\ref{fig:k102} to \ref{fig:k872}). All labeled features are at the $>5\sigma$ significance level. However, in sources like KISSR1494 (Figure \ref{fig:k1494}), the lowest contour is chosen to be $\approx2\sigma$ to primarily highlight the diffuse emission (around component J1).} KISSR967, which was detected at the $\sim5\sigma$ level with a total flux density of $\sim160~\mu$Jy at 1.5~GHz in January 2019 is not detected $\sim3.7$ yrs later. For this source image, the r.m.s. noise is higher in epoch-II compared to epoch-I by a factor of $1.2-1.4$. However, the drop in core flux density (assuming $3\sigma$ as an upper limit to core flux density in epoch-II) between the two epochs is $\sim27$\%. The non-detection of emission from KISSR967 could therefore be related to intrinsic source variability. 

Table~\ref{tab:prop} lists the `core' flux density changes that occurred within the time interval between the two epochs. The largest changes in `core' flux density are observed in the case of KISSR872 ($39\pm6$\%) and KISSR102 ($29\pm7$\%) in a time period of 3.66 yr and 3.69 yr, respectively (see Section~\ref{sec:change}). Interestingly, these 2 LINER galaxies are also the 2 sources showing the presence of superluminal jet motion. For the remaining sources, the `core' flux density changes are not significant.
We discuss changes in jet morphology and spectral indices between the two observation epochs, below. 

\subsection{VLBI `cores' and Gaia Optical Positions}\label{sec:gaia}
Our assumption of the brightest feature being the radio `core' seems largely to be validated in these KISSR galaxies. With the exception of KISSR872 and KISSR618, the highest intensity features are also the `stationary' features. We have included the Gaia DR3 optical positions of the centres of the host galaxies in Figures~\ref{fig:k102}-\ref{fig:k872}. With the exception of KISSR102, the VLBI `cores' in all sources are offset by 21 to 169~mas (see Table~\ref{tab:prop}). However, this offset may be understood as large astrometric uncertainities in the optical positions. For instance, the $astrometric\_excess\_noise$ parameter, which is the extra noise that is needed per observation to explain the scatter of residuals in the astrometric solutions \citep{Lindegren2021}, is large for the KISSR galaxies ($\sim5-33$~mas). Another goodness-of-fit parameter is the renormalised unit weight error \citep[ruwe;][]{Lindegren2018}. This parameter is expected to be close to 1.0 for well behaved sources, and has been estimated in Gaia DR3 only for KISSR102 (ruwe=1.87), KISSR872 (ruwe=3.23) and KISSR1494 (ruwe=1.74), suggesting a relatively poor astrometric fit. Consistent with this suggestion, we find that the largest offsets between the VLBI `core' and the Gaia positions are in KISSR434, KISSR1154 and KISSR1494, respectively, which are inclined spiral galaxies with prominent spiral arms and dust lanes. The presence of the extended light from the galactic bulge, dust lanes, circumnuclear star clusters, can shift the optical centroid away from the SMBH. The weakness and the variability of the optical AGN emission can be an additional contributor to astrometric uncertainities in the KISSR galaxies.

\subsection{Global Properties of the KISSR Galaxies}\label{sec:global}
We list some of the global properties of the sample sources in Table~\ref{tab:global}. The 6~GHz radio luminosity, L$_6$, was estimated by extrapolating the integrated flux density at 1.4~GHz from the VLA FIRST survey with a global spectral index of $-0.7$. All the sample sources satisfy the `radio-quiet' classification as defined by \citet{Kellermann2016}, i.e., have $~10^{21} \lesssim L_6 \lesssim 10^{23.2}$~W~Hz$^{-1}$. The time-averaged kinetic jet power of the radio outflows in the KISSR sources was estimated following the relations used for `radio-loud' AGN \citep[see][]{Punsly11}. We derived the flux density at 151~MHz, $F_{151}$, that is needed for the calculation, using the 1.4~GHz flux density from the kpc-scale FIRST images and a jet/lobe spectral index of $-0.7$. The star-formation rate (SFR) is derived from the H$\alpha$ narrow line luminosity and the \citet{Kennicutt98} relation, while the black hole mass is based on the M$-\sigma_\star$ relation \citep{McConnell13}. The bolometric luminosity, L$_\mathrm{bol}$, is based on the [O III] line luminosity and the scaling relation from \citet{Heckman04}. We see that the L$_\mathrm{bol}$ for the sources range between $10^{43}-10^{44}$~erg~s$^{-1}$; the Seyfert galaxies in this sample therefore have low luminosities similar to the LINER galaxies which typically have L$_\mathrm{bol}\leq10^{43}$~erg~s$^{-1}$ \citep[e.g.,][]{Younes2012, Spinoglio2024}.

\subsection{VLBI Features between Epochs: Changes in Flux density, Spectral index \& Jet Directions}\label{sec:change}
Overall in the second epoch of observations, we find a persistence of the weak radio features that were observed in the first epoch observations, attesting to their reality. {Except for KISSR102, we did not detect flat or inverted spectral index `cores' that are consistent with synchrotron self-absorbed bases of jets.} Instead, we had identified the `cores' to be the brightest centrally peaked features. These second epoch phase-referenced observations have {largely} confirmed that these `cores' are indeed `stationary' features, validating our assumptions. The `core' position remains the same to within 1 mas in a majority of the sources; the peak intensity positions primarily shift by 2.0~mas and 2.9~mas in KISSR618 and KISSR872, respectively. These 2 sources also show the most variation in its jet base over the 2 epochs. However, considering the limited astrometric precision and the availability of only two epochs of data, the `cores' may ideally be regarded as `candidate cores'. In principle, extremely high-precision differential astrometry \citep[e.g.][]{Cheng2023} is required to confirm their stationarity.

The candidate cores in several sources, viz. KISSR434, KISSR618, KISSR872, and KISSR1494, show associated diffuse emission in a transverse direction. For KISSR618, the broadening of the jet occurs 10~mas from the `core'. The width of these jet bases are respectively, 2.4, 1.9, 1.9 and 1.9 times the major axes of their synthesized beams. It is not yet clear if these structures are similar to the broad base observed in say, the RadioAstron image of 3C84 \citep{Giovannini2018}. Their presence will need to confirmed in future observations of the sample. We discuss this more in Section~\ref{discussion}.

It is unclear whether there is a jet in the epoch-II image of KISSR1154. The presence of a weak jet was indicated in this source in the epoch-I data \citep{Kharb2021}, attesting to the rapid changes in the jet structures in these sources. Epoch-I observations of KISSR1494 had detected only a candidate core \citep{Kharb2015}; however, a jet is clearly detected in epoch-II observations, 9.2 years later. As the change in sensitivity is not significant between the two epochs (see Table~\ref{tab:global}) the jet features appear to have significantly brightened over the last 9.2 years. 

Significant `core' flux density change is observed in KISSR102 ($29\pm7$\%) and KISSR872 ($39\pm6$\%). These 2 sources also exhibit superluminal motion in their jets (see Table~\ref{tab:prop}). The `core' flux density change is not significant in the remaining sources. The flux density change error has been estimated using the rules of error propagation: $100 \times \sqrt{ \left( \frac{S_2\,\sigma_{S1}}{S_1^2} \right)^{2} + \left( \frac{\sigma_{S2}}{S_1} \right)^{2} }$, where $\sigma_{S1}=\sqrt{\mathrm{rms}_1^2 + (0.05\,S_1)^2}$ and $\sigma_{S2}=\sqrt{\mathrm{rms}_2^2 + (0.05\,S_2)^2}$, `rms' being the r.m.s. noise levels in the respective images and the flux-density-scale error for the VLBA assumed to be 5\%. 

Of the 3 sources that were observed at 5~GHz in epoch-II, only 2 sources, viz. KISSR102 and KISSR872, were detected at 5~GHz. KISSR434 was not detected at 5~GHz in this second epoch; in the previous epoch, this source had exhibited steep spectrum emission ($\alpha = -0.9$ to $-1.3$) in its candidate core region \citep{Kharb2019}. The `core' spectral index in KISSR102 changed from $+0.64\pm0.08$ in epoch-I to $+0.91\pm0.13$ in epoch-II, 3.7 years later. This source is discussed in greater detail in Section~\ref{sec:k102}. The flattening of the $1.5-5$~GHz spectral index from $-0.71\pm0.26$ to $-0.31\pm0.19$ for the core-jet region in KISSR872 has already been discussed in \citet{Kharb2024}. Therefore, both the jet structures and jet spectral indices vary between epochs. 

We had fitted a precessing jet model to {the curved jet} KISSR434 in \citet{Kharb2019}. {The curved jet in KISSR434 continues to remain curved 4 years later.} The persistence of the curvature disfavors the jet nozzle precession model but supports the idea that the jet plasma is flowing along a curved path. In KISSR618, the pc-scale jet has changed its direction of propagation by nearly 30$\degr$. There also appears to be flaring in the inner jet around 10~mas from the core. However, overall there remains some {extended} emission in the previously observed jet direction. 

\subsection{Estimating Jet Speeds in KISSR Sources}
A comparison of the jet knot positions in 2 of the LINER galaxies, viz., KISSR872 and KISSR102, indicates the presence of relativistic jet motion. As far as we are aware, this makes KISSR872 and KISSR102 the only 2 radio-quiet LINER galaxies to reveal relativistic or superluminal motion in their pc-scale jets. A jet knot is seen to move at a speed of $1.65\pm0.57$c in $\sim4$ years in KISSR872 \citep{Kharb2024}. KISSR102 showed the presence of two compact radio cores in the image from December 26, 2018 \citep{Kharb2020} but now reveals superluminal jet motion (see Section~\ref{sec:k102}). {Recently, \citet{Wang2025} have detected superluminal jet motion ($\beta_{app}=1.5 - 3.6$) in the `radio-quiet' narrow-line Seyfert 1 galaxy, Mrk110.}

So far, we have closely examined 2 of the 7 sources with the simplest core-jet structures. More sophisticated methods like those discussed in Section~\ref{sec:k102} will need to be employed for the other sources that show jet features that are not compact making them difficult to identify and trace (e.g., KISSR434). The jet components observed in epoch-I 7.75 years ago in KISSR1219, appear to have moved away. If indeed, J1 and J2 (see Figure~\ref{fig:k1219}) are the same components as seen in epoch-I for KISSR1219, the apparent jet speed is 2.42c and 0.88c, respectively. VLBI monitoring is therefore required to confirm the jet speeds in KISSR1219 and other sources. 

\begin{deluxetable*}{lcccclcc}\label{tab:global}
\tabletypesize{\scriptsize}
\tablewidth{0pt} 
\tablecaption{Global properties}
\tablehead{\colhead{Source} & \colhead{Black hole} & \colhead{$\mathrm{L_{bol}}$} & \colhead{$S^{\mathrm{total}}_{\mathrm{FIRST}}$} & \colhead{L$_6$} & \colhead{Eddington} & \colhead{$\mathrm{Q_{jet}}$} & \colhead{SFR} \\
\colhead{name} & \colhead{mass (M$_\sun$)} & \colhead{erg~s$^{-1}$} & \colhead{mJy} & \colhead{W~Hz$^{-1}$} & \colhead{ratio}  & \colhead{erg~s$^{-1}$}  & \colhead{M$_\sun$~yr$^{-1}$}}
\colnumbers
\startdata
KISSR102 & $1.7\times10^9$ & $6.40\times10^{43}$ &$11.3\pm0.3$ &$4.01\times10^{22}$ & $3.0\times10^{-4}$ & $8.9\times10^{41}$ & $0.30\pm0.01$\\
KISSR434 & $1.3\times10^8$ & $1.91\times10^{44}$ &$6.0\pm0.3$ &$1.98\times10^{22}$ & 0.012  &$4.9\times10^{41}$ & $0.7\pm0.1$\\ 
KISSR618 & $4.6\times10^7$ &$1.09\times10^{44}$  &$2.5\pm0.3$ &$1.08\times10^{22}$ & 0.018  &$2.9\times10^{41}$ & $0.36\pm0.03$\\
KISSR872 & $4.4\times10^7$ &$2.26\times10^{44}$  &$5.2\pm0.3$ &$2.96\times10^{22}$ & 0.040  &$6.7\times10^{41}$ & $0.93\pm0.08$\\
KISSR967 & $1.4\times10^8$ &$2.30\times10^{43}$  &$2.4\pm0.2$ &$1.70\times10^{22}$ & 0.0013 &$4.1\times10^{41}$ & $1.36\pm0.20$\\
KISSR1154 & $3.6\times10^7$&$8.70\times10^{43}$  &$3.5\pm0.8$ &$1.48\times10^{22}$ & 0.019  &$3.8\times10^{41}$ & $0.13\pm0.04$\\
KISSR1219 & $2.1\times10^7$& $6.19\times10^{43}$ &$5.6\pm0.3$ &$6.11\times10^{21}$ & 0.020  &$1.9\times10^{41}$ & $0.40\pm0.07$\\  
KISSR1494 & $1.4\times10^8$& $4.20\times10^{44}$ &$23.0\pm0.3$ &$6.04\times10^{22}$ & 0.020   &$1.3\times10^{42}$ & $1.70\pm0.09$
\enddata
\tablecomments{{Column 1: source name.} Columns 2 and 3: black hole estimates using the M$-\sigma_\star$ relation and bolometric luminosity, respectively (see Section~\ref{sec:global} for details). Column 4: Total flux density in the VLA FIRST image with errors obtained from the Gaussian-fitting {\tt AIPS} task {\tt JMFIT}. Column 5: 6~GHz luminosity extrapolated from VLA FIRST 1.4~GHz data using $\alpha=-0.7$; these are consistent with the sources being `radio-quiet' as defined by \citet{Kellermann2016}. Columns 6 and 7: Eddington ratios and time-averaged jet kinetic power; the latter was derived using the VLA FIRST total flux densities and $\alpha=-0.7$ to convert to 151~MHz flux densities and the relations in \citet{Punsly11}. Column 8: star formation rate derived from the narrow H$\alpha$ line luminosity (Section~\ref{sec:global}).}
\end{deluxetable*}

\subsubsection{A Relativistic Jet in KISSR102}\label{sec:k102}
We used the Cross-Entropy (CE) global optimization technique \citep{Rubinstein_1997} to model our 4 VLBI images of KISSR102. The CE optimization was originally adapted by \citet{Caproni_et_al_2011} to deal with the problem of discretizing interferometric radio images in terms of two-dimensional elliptical Gaussian components, having been applied to other jetted radio sources in former works \citep{2014MNRAS.441..187C, 2017ApJ...851L..39C, 2021MNRAS.509.1646S, 2024ApJ...965....9N}. Each elliptical Gaussian has 6 free parameters to be determined: peak intensity, $I_0$, two-dimensional peak position ($x_0, y_0$), with the coordinates $x$ and $y$ oriented respectively to right ascension and declination directions, semi-major axis, $a$, eccentricity, $\epsilon=\sqrt{1-(b/a)^2}$, where $b$ is the semi-minor axis, and the position angle of the major axis, $\psi$, measured positively from west to north. 

In short, CE optimization tests each tentative solution involving the combination of $N_\mathrm{c}$ elliptical Gaussian components against a predefined merit function. It considers the twenty solutions with the lowest merit function values to build the next sample of tentative solutions (e.g., see \citealt{2014MNRAS.441..187C} for more details). The CE optimization is halted when either the maximum number of iterations is reached or the r.m.s. value of the associated residual image falls below the nominal r.m.s. value provided in Table~\ref{tab:results}. This process is repeated twice for each interferometric image, selecting the CE optimization that best minimizes the merit function for determining the structural parameters of the Gaussian components and their respective uncertainties (see \citealt{Caproni_et_al_2011} for more details).

Using the criteria proposed by \citet{2014MNRAS.441..187C}, we determined the optimal number of elliptical Gaussian components in each of the 4 VLBI images of KISSR102. For the two radio maps at 4.98 GHz, one single Gaussian component is enough to represent the brightness distribution of KISSR102, while two Gaussian components were necessary to provide a fair description of the two 1.54 GHz images. We show in Figure~\ref{fig:KISSR102_FluxDens} the time behavior of the flux density of the Gaussian components A and B for the 2 epochs analyzed in this work. Both components show changes in their flux densities between $\sim 2019$ and 2022.7. Component A becomes slightly fainter at 4.98 GHz and brighter at 1.54 GHz, while component B decreases its flux density at 1.54 GHz. Comparing the flux densities of the component A at 1.54 and 4.98 GHz, we obtained $\alpha\sim1.4$ in 2019 and $\alpha\sim1.1$ in 2022.7, implying that A is optically thick in both epochs. These findings are in agreement with the spectral index map of KISSR102 shown in the bottom panel of Figure~\ref{fig:k102}, suggesting that A is likely the core region of KISSR102 and B a jet component.

In the top panel of Figure~\ref{fig:KISSR102_RADEC_rxt}, we show the offsets in right ascension and declination of component B relative to core A at 1.54 GHz, while its core-component separation is presented in the bottom panel. Between $\sim 2019$ and 2022.7, component B has receded from A at a mean position angle around $-42\degr$ and with a proper motion of $0.24\pm0.10$ mas yr$^{-1}$, which translates to an apparent speed of $(1.05\pm0.45)c$. Superluminal motions are commonly reported in blazars but not in LINERS, for which KISSR872 is the only documented case in the literature \citep{Kharb2024}. 

Interestingly, the mean position angle of the component B ($-42\degr$) agrees with the position angle of the secondary optical nucleus N2 seen in SDSS images of KISSR102 (see Figure 1 in \citealt{Kharb2020}). It suggests an alternative interpretation for N2 in which it would be a kpc-scale jet component ejected from N1 $\sim 5000$ years ago (considering that N2 has the same proper motion as the pc-scale jet component B). This possibility could also explain the poor modeling obtained by \citet{Kharb2020} after assuming N2 as a point source since jet components are expected to expand as they recede from the core region. This interpretation also suggests a possible interaction between the jet and the interstellar medium (`AGN feedback'), which would be in full agreement with optical forbidden emission lines being excited by shocks in the case of KISSR102 \citep{Kharb2020}. Higher resolution VLA images will be required to confirm this hypothesis.

\begin{deluxetable*}{lccccl}\label{tab:prop}
\tabletypesize{\scriptsize}
\tablewidth{0pt}
\tablecaption{Global VLBI properties}
\tablehead{
  \colhead{Source} & \colhead{Offset} & \colhead{Gaia Noise} & \colhead{Flux density} & \colhead{Time} & \colhead{Qualitative findings and a comparison}\\
  \colhead{name}   & \colhead{mas}    & \colhead{mas} & \colhead{change (\%)} & \colhead{diff. (yr)} & \colhead{with previous epoch observations} } 
\colnumbers
\startdata
KISSR102  & 1.52 & 8.08 & $29\pm7$   & 3.69 & Superluminal jet motion with $v=1.05\pm0.45$c\\
KISSR434  & 169.24 & 20.28 & $14\pm14$   & 4.85 & Broad jet \\
KISSR618  & 22.25  & 12.35 & $12\pm14$   & 3.66 & Broad jet with jet flaring; Jet  direction change \\
KISSR872  & 38.38\tablenotemark{a}  & 7.14  & $39\pm6$ & 3.66 & Broad jet; Changes b/w epochs; Superluminal jet motion with $v=1.65\pm0.57$c\\
KISSR967  &\nodata& 23.31 & $>27$\tablenotemark{b} & 3.67 & No detection in Epoch-II; Weak core with eastern extension detected in Epoch-I\\
KISSR1154 & 61.89 & 32.86 & $19\pm20$   & 3.72 & No jet detection in Epoch-II; North-eastern jet component suggested in Epoch-I \\
KISSR1219 & 21.23 & 5.78 & $3\pm20$  & 7.75 & Possible broad jet needing confirmation \\
KISSR1494 & 40.00 & 5.07 & $2\pm30$  & 9.18 & Broad jet, No jet detection in Epoch-I\\
\enddata
\tablenotetext{a}{Epoch-I image with a more clearly defined `core' was considered.}
\tablenotetext{b}{KISSR967 was not detected in epoch-II.}
\tablecomments{Column 1: source name. Column~2: offset between VLBI `core' and Gaia DR3 host galaxy position in milli-arcseconds. Column~3: the Gaia $astrometric\_excess\_noise$ parameter in milli-arcseconds (see Section~\ref{sec:gaia}). Column 4: fractional percentage change in the VLBA `core' flux density between 2 epochs with errors (Section~\ref{sec:change}). Column 5: time difference between the 2 VLBA epochs in years.}
\end{deluxetable*}

The bulk speed of the jet in terms of $c$, $\beta$, and the viewing angle of the jet, $\phi$, can be determined univocally from the apparent speed of the jet, $\beta_\mathrm{app}$, defined as
\begin{equation}
\beta_\mathrm{app} = \frac{\beta \sin \phi}{(1-\beta \cos \phi)},
\label{eq:betaapp}
\end{equation}
and the measured flux densities at a frequency $\nu$ for the jet, $S_{\nu,\mathrm{j}}$, and counterjet, $S_{\nu,\mathrm{cj}}$. 
Assuming the flux densities of the jet and counter-jet in the source's reference frame are equal, any difference between $S_{\nu,\mathrm{j}}$ and $S_{\nu,\mathrm{cj}}$ is due to the relativistic Doppler boosting effect, so that for a continuous jet (e.g., \citealt{1985ApJ...295..358L})
\begin{equation}
\frac{S_{\nu,\mathrm{j}}}{S_{\nu,\mathrm{cj}}}= \left(\frac{1+\beta \cos \phi}{1-\beta \cos \phi}\right)^{2-\alpha}.
\label{eq:Snujcjratio}
\end{equation}
Thus, using $\beta_\mathrm{app}$ ($=1.05c$) of the jet component B estimated in this work, together with $S_{\nu,\mathrm{j}}/ S_{\nu,\mathrm{cj}}\approx 20$ and $\alpha=-1.6$ \citep{Kharb2020}, we derived $\beta\approx 0.75c$ and $\phi\approx 58\degr$ for the pc-scale jet of KISSR102. 

These values allow a rough estimation of the maximum projected distance from A that the counter-jet has reached in the last 5000 years. Assuming that jet components are ejected from the core at $0.75c$ at a viewing angle of $58\degr$, maintaining both values during their displacements, the counter-jet must have extended to an angular distance of about 0\farcs5 from A. It corresponds to twice the pixel size of the $i$-band image of KISSR102 shown in the left panel of Figure 4 in \citet{Kharb2020}. Therefore, a possible interaction between the counter-jet and the interstellar medium that could be detected at optical wavelengths would be barely resolved in the referred $i$-band image of KISSR102. However, the excess seen in the residual map close to the location of N1 \citep[right panel in Figure 4 of][]{Kharb2020} might indicate underlying diffuse emission that could be due to the counterjet itself. 

\subsection{Extended Radio Emission in the VLA Images}\label{sec:vla}
While jets are not observed in the VLA images, extended (extranuclear) emission is detected in all 4 sources (Figure~\ref{fig:fig10}), the morphology of which is typically core-lobe-like\footnote{The diffuse emission shows a directionality.} (in KISSR434, KISSR872, and KISSR967) or core-halo-like\footnote{The diffuse emission is symmetrically distributed around the core.} (in KISSR618). The extension in KISSR967 is however along the direction of the second optical nucleus observed in this galaxy. 

The in-band spectral indices shown in Figure~\ref{fig:fig10} reveal an average spectral index ranging from $-0.7\pm0.3$ to $-0.90\pm0.06$, consistent with synchrotron emission. {Spectral index gradients are observed in all the 4 sources. These gradients may be hinting at the directions of unresolved jetted structures, with the nuclei having a flatter spectrum and the jets having a steeper spectrum. We find that the VLBI jet directions are aligned with the spectral index gradients in all the 4 sources. This strongly supports an AGN-jet-related origin for the VLA-scale emission.}

Moreover, if we assume that all of the observed radio luminosity is attributable to star formation, an SFR can be derived using the relations in \citet[][Equation 21]{Condon1992}. We have derived these SFRs for stellar masses $\ge5$~M$_{\sun}$ (SFR1) and for stellar masses M$\ge0.1$~M$_{\sun}$ (SFR2) in Table~\ref{tab:sfr}. SFR2 is $\approx5.5$ times SFR1 \citep{Condon2002}. For these calculations, we have subtracted the peak flux density of the radio emission which is associated with the AGN itself, from the total radio flux density. The average spectral index used for the estimation is noted in Table~\ref{tab:sfr}. A comparison with the SFR derived from the H$\alpha$ emission lines in Table~\ref{tab:global} shows that the SFR (M$\ge5$~M$_{\sun}$) derived from the radio emission is between 5 to 25 times larger (these extreme values are for KISSR872 and KISSR618, respectively). For stellar masses M$\ge0.1$~M$_{\sun}$, these values differ by factors ranging from 28 to 136 (for KISSR872 and KISSR618, respectively) with the radio SFR therefore being $2-3$ orders higher than the H$\alpha$ derived SFR. We conclude that the extended radio emission observed in the VLA images is likely to be AGN-related rather than stellar or supernovae-related. 

Polarization is detected at the edge of the lobe in KISSR872 (Figure~\ref{fig:fig11}). The fractional polarization is $28\pm9\%$ at a polarization angle of $14\pm8\degr$. None of the other 3 sources reveal any polarization. Interestingly, the detected polarization in KISSR872 is in the direction of the secondary nucleus in this interacting galaxy. The SDSS image of KISSR872 reveals not only the tidal tail to the south-east of the primary nucleus but also a faint tail towards the north, in between the two optical nuclei, in fact just north of the detected polarization. We discuss a possible interpretation in Section~\ref{discussion}.

Ascribing the extended radio emission seen in the VLA images for the 4 KISSR sources {to the AGN} implies that already at the $\sim1\arcsec$ scales (typical FWHM of the VLA synthesised beam at this frequency) which corresponds to spatial scales of $1.0-1.5$ kpc-scale for these sources, the jets in these sources have lost collimation or `flared' into radio lobes. As the largest `jets' that we have observed in our VLBA observations \citep[e.g.,][]{Kharb2019,Kharb2021} were of $100-200$~parsec extents, the jet decollimation regions are between $\sim200$~parsec (which is a limit imposed by the field of view of the VLBA) and $\sim1$~kpc. Higher resolution observations, say with the eMERLIN array at 1.4~GHz are required to probe these regions of jet decollimation. In terms of the relevant scales, a few hundred parsecs is the radius where there is indeed a sharp transition in the physical characteristics of the host galaxy ISM \citep[e.g.,][]{Launhardt2002}. {Changes in pressure equilibrium between the jet and the ambient medium have indeed been suggested to produce jet `flaring' and deceleration on scales of a few kpc in FRI radio galaxy jets by \citet{Laing2002}. A similar scenario may hold in the relatively lower speed jets in Seyfert and LINER galaxies on shorter distances from their nuclei.}

\begin{deluxetable*}{lcccclcc}\label{tab:sfr}
\tabletypesize{\small}
\tablewidth{0pt} 
\tablecaption{Results from the VLA 1.5~GHz A-array observations\label{tab:sfr}}
\tablehead{
\colhead{Source} & \colhead{Total flux} & \colhead{Peak flux} & \colhead{Extd. flux}& \colhead{Average} & \colhead{Luminosity} & \colhead{SFR1} & \colhead{SFR2} \\
\colhead{name} & \colhead{density (mJy)} & \colhead{density (mJy)} & \colhead{density (mJy)}& \colhead{in-band $\alpha$}&\colhead{W~Hz$^{-1}$} & \colhead{M$_\sun$~yr$^{-1}$} & \colhead{M$_\sun$~yr$^{-1}$}} 
\colnumbers
\startdata
{KISSR434}& 4.168 & 2.477 & 1.691 & $-0.90\pm0.06$ & 1.55E+22 & 4.43 & 24.37 \\
{KISSR618}& 3.897 & 0.946 & 2.952 & $-0.8\pm0.1$ & 3.55E+22 & 8.90 & 48.96\\
{KISSR872}& 5.716 & 4.558 & 1.159 & $-0.7\pm0.3$ & 1.83E+22 & 4.76 & 26.18 \\
{KISSR967}& 3.492 & 0.916 & 2.576 & $-0.80\pm0.06$ & 5.07E+22 & 13.82 & 75.99 \\
\enddata
\tablecomments{{Column 1: source name.} Columns 2, 3, 4: the total, peak and extended (total$-$core) flux density at 1.5~GHz in the VLA images, respectively. Column 5: the average in-band spectral index value. Column 6: the 1.5~GHz total radio luminosity. Columns 7, 8: the derived star formation rates with stellar masses $\ge5$M$_{\sun}$ and $\ge0.1$M$_{\sun}$, respectively (see Section~\ref{sec:vla}.)}
\end{deluxetable*}

\section{Discussion}\label{discussion}
Second epoch VLBI images of 8 KISSR galaxies with double-peaked or asymmetric lines in their optical spectra reveal core-jet structures with superluminal jet motions in at least 2 of them, viz., KISSR102 and KISSR872 \citep{Kharb2024}. One of these 8 sources, viz. KISSR967, which was detected in epoch-I is not detected in epoch-II. Distinct and compact jet knots that can be identified and traced in both epochs are typically not observed in these radio-faint jets, making the task of speed estimation harder. 

A change in jet propagation direction is {possibly} observed in KISSR618. The jet curvature identified in KISSR434 in epoch-I remains that way 4 years later in epoch-II. A core-jet structure is identified in KISSR1494 in epoch-II which was not {detected} in epoch-I. Overall, the presence of a jet is evident in the VLBI images of all the 7 detected KISSR galaxies. Moreover, fast variability in both flux density and pc-scale radio morphology is evident between two epochs of VLBI data. 
The average candidate core flux density change for the 2 sources where the change is significant, viz. KISSR102 and KISSR872, is $\approx30$\%. Similar levels of flux density variability ($\sim20-30$\%) have been observed in other radio-quiet AGN like NGC5548 \citep{Wrobel2000}, and NGC4051 \citep{Jones2011}. 

In at least 4 sources, viz. KISSR434, KISSR618, KISSR872, and KISSR1494, but perhaps also in KISR1219, the 1.5~GHz `cores' show diffuse emission extended in a transverse direction resembling a broad jet base. Pending their confirmation in future VLBA observations, we discuss some tentative suggestions for their origin. The broad base observed in the RadioAstron image of 3C84 \citep{Giovannini2018} was suggested to arise from the accretion disk. The results from the KISSR sources may be hinting at the existence of jet stratification with either a spine + sheath structure or a jet + (radio-emitting) wind structure on pc-scales \citep[e.g.,][]{Nevin2016, Silpa2023, Kharb2023Galax, Ghosh2025}. A jet+wind structure is also consistent with the flattening of the average core-jet spectral index in KISSR872 \citep{Kharb2024} over a period of 3.7 yrs (see Figure~\ref{fig:k872}). The spectral index of KISSR102 became more inverted in epoch-II. This may be consistent with the ejection of a steep-spectrum pc-scale jet component leaving behind a more compact inverted-spectrum core. 

Changes in the direction of jet propagation have been observed in several blazars in the MOJAVE sample \citep{Lister2018}; these could imply precession of the jet nozzle. A nearly 30$\degr$ change is observed in the case of KISSR618. Although, the persistence of jet curvature in KISSR434 disfavors jet nozzle precession in that source but rather supports the idea that the jet plasma flow is along a curved path. Such a behaviour has indeed been observed in the blazar 1156+295 \citep{Zhao2011} and the quasar 0836+710 \citep{Hummel1992}. 

{While some of the jet behavior in the KISSR sources is reminiscent of that observed in `radio-loud' AGN, there are subtle differences as well. The cores and jet knots in these `radio-quiet' KISSR sources lack the compactness and brightness typically observed in the pc-scale features of `radio-loud' AGN. This might be indicative of intrinsic differences in the jet composition and/or magnetic field strengths between `radio-loud' and `radio-quiet' AGN \citep[e.g.,][]{Gulati2025}.}

The VLBI images of the KISSR galaxies are consistent with the 2D slab-jet simulations of \citet{Saxton2005}, which were used to explain the VLBI images of the compact steep spectrum (CSS) quasar 3C48. The simulation results of \cite{Saxton2005} with low density jets propagating in an inhomogeneous medium with a small filling factor of warm, dense clouds, could match what we observe in the VLBI images of the KISSR sources. These outflows are supposed to create bubbles later which are observed in several Seyfert and LINER galaxies \citep[e.g.,][]{Kharb06, Sebastian2019, Ghosh2025}, although not seen in our sources on the spatial scales explored. Apart from the caveats discussed by \cite{Saxton2005}, which include the two-dimensionality of the simulations and the non-relativistic flows considered, another caveat is that the jet powers considered in \cite{Saxton2005} fall in the range of $10^{45}-10^{46}$~ergs~s$^{-1}$, whereas the jet powers estimated for the KISSR sources fall in the range of $10^{41}-10^{42}$~ergs~s$^{-1}$.

The jets in the KISSR galaxies are likely to be `light' \citep[see the discussion on a jet-to-ambient-medium density ratio, $\eta$, of $10^{-2}$ in the KISSR434 jet from ram pressure bending and jet kinetic power arguments in][]{Kharb2019} and `fast' \citep[see the relativistic jet motion in KISSR872 in][]{Kharb2024}. Fast but light jets are likely to have low kinetic jet powers, consistent with what is obtained for the KISSR sources. As these jets traverse an inhomogeneous medium with a small filling factor, whose characteristics have not been probed so far, they may break or become filamentary on scales of a few 100 parsecs.  

{No jetted structures are observed in the VLA images of the 4 observed galaxies.} However, S-shaped lobe-like structures may be present in KISSR434 and KISSR872, which need higher resolution observations to confirm. The kpc-scale spectral index gradients observed {in all 4 sources lies} along the VLBI jet directions. This {strongly} supports the suggestion that the VLA-scale radio emission is predominantly AGN-jet-related. {This is consistent with the finding that the SFR derived from the radio luminosity is a factor or a few (for M$\ge5$~M$_{\sun}$) to several orders of magnitude (for M$\ge0.1$~M$_{\sun}$) larger than that derived from the H$\alpha$ emission lines.} Considering both the VLBI and VLA A-array images, the jets likely lose collimation between $\sim$200 parsec and $\sim$1 kpc (the lobe-like structures in KISSR434 and KISSR872 extend up to 4 and 5 kpc, respectively). 

Kpc-scale polarization is only detected in the LINER galaxy KISSR872. Interestingly, the region of high fractional polarization is in the same direction as the second optical {nucleus} present in this interacting galaxy. The inferred magnetic field in this optically thin region is perpendicular to the polarization vectors \citep{Pacholczyk1970}, as well as this northern fainter tidal tail. This is suggestive of compressed gas and magnetic fields as the gas gets tidally pulled from the secondary nucleus by the primary nucleus. KISSR872 resembles the interacting galaxy NGC6240, which is in a late-stage merger, and possesses a bridge of hot, turbulent gas in between its two optical nuclei \citep{Treister2020,Munoz2025}.

\section{Summary and Conclusions}
We have presented the second epoch phase-referenced VLBA images of 8 KISSR galaxies at 1.5~GHz and 3 of them being observed at 5~GHz as well. We summarise the primary findings below.

\begin{enumerate}
\item We detect {emission in} 7 out of 8 KISSR galaxies at 1.5~GHz in the second epoch of VLBI observations; {KISSR967 was not detected in this epoch}. We detect {emission in} 2 out of 3 sources at 5~GHz in epoch-II observations; {KISSR434 was not detected at 5~GHz in this epoch}. Changes are observed in the flux density, spectral index, jet features, and jet propagation directions in these galaxies. In one source, viz. KISSR434, the jet curvature remains after a period of 4 years, {while in another source, viz. KISSR618, the jet direction may have changed by nearly 30$\degr$}. Such a behaviour in pc-scale jets is also observed in the VLBI observations of `radio-loud' AGN.
\item Superluminal jet motion is observed in the jets of 2 LINER galaxies, viz., KISSR102 and KISSR872 \citep{Kharb2024}. The determination of jet speeds in other sources is made difficult by the presence of diffuse jet components which are hard to identify between epochs. Multi-epoch VLBI observations are necessary to study the jets in Seyfert and LINER galaxies. 
\item Sub-structures are observed in the {cores and jets} of several sources possibly indicating jet stratification on pc-scales. Relatively broad jets might be consistent with `jet+wind' structures or `spine+sheath' structures. This {might} be a result of jet-medium interaction as suggested in the jet simulations of \citet{Saxton2005}. Faster lighter jets moving in inhomogeneous media with small filling factors {could} explain these VLBI results. 
\item The VLA images of 4 KISSR sources reveal core-halo or core-lobe structures. The alignment of the pc-scale jets with the kpc-scale spectral index gradients from flat-to-steep is consistent with an AGN-jet-related origin of the VLA emission. A large mismatch between the SFR ($5-25$ for stellar masses M$\ge5$M$_\sun$) derived from the radio emission and H$\alpha$ line emission also supports an AGN origin for the radio emission. A comparison of the VLBA and VLA images suggests that the jets in the KISSR sources lose collimation on spatial scales between 200 parsec and 1 kpc. 
\item Signatures of jet-mode AGN feedback on kpc-scales may be present in KISSR102. However, sub-kpc-scale images are needed to confirm the presence of a jet on these scales. The jet-medium interaction has also been invoked to explain the emission line analysis in \citet{Kharb2021} as well as the presence of double peaks and asymmetries in the emission lines in KISSR sources.
\end{enumerate}

Overall, our two-epoch VLBI study of 8 KISSR Seyfert and LINER galaxies strongly supports the presence of $\sim$100~pc-scale jets in these `radio-quiet' AGN. The behavior of these jets is similar to the pc-scale jets observed in `radio-loud' AGN, although they are highly scaled down in terms of total radio power. {There are, however, subtle differences as well. The cores and jet knots in these `radio-quiet' sources lack the compactness and brightness typically observed in the pc-scale features of `radio-loud' AGN. This might be indicative of intrinsic differences in the jet composition and/or magnetic field strengths between `radio-loud' AGN and `radio-quiet' AGN.}

\section{Acknowledgments}
{We thank the anonymous referee for their comments that have improved this manuscript significantly.}
PK and SG acknowledge the support of the Department of Atomic Energy, Government of India, under the project 12-R\&D-TFR-5.02-0700. PK acknowledges the useful suggestions from Silpa Sasikumar on EVLA data analysis. MD gratefully acknowledges the support of the Department of Science and Technology (DST) grant DST/WIDUSHIA/PM/PM/2023/25(G) for this research. The National Radio Astronomy Observatory is a facility of the National Science Foundation operated under cooperative agreement by Associated Universities, Inc. The VLBA data were calibrated using NRAO's ``VLBA data calibration pipeline'' in AIPS. {This work has made use of data from the European Space Agency (ESA) mission Gaia.}

\vspace{5mm}
\facilities{VLBA, VLA, Gaia}
\software{AIPS \citep{Greisen2003}, CASA \citep{McMullin2007}}

\bibliography{ms}{}
\bibliographystyle{aasjournal}
\end{document}